\documentclass[aps,prd,twocolumn,groupedaddress, showkeys, showpacs]{revtex4}
\usepackage{graphicx,subfigure,enumerate,pstricks,latexsym}
\usepackage[english]{babel}

\providecommand{\dif}{\mathrm{d}}
\def\beq{\begin{equation}}
\def\eeq{\end{equation}}
\def\bea{\begin{eqnarray}}
\def\eea{\end{eqnarray}}

\def\nn{\nonumber}

\newcommand{\nnd}{\discretionary{--}{--}{--}}
\newcommand{\Schw}{Schwarzschild}
\newcommand{\dS}{de~Sitter}

\def\linka{ --- }

\def\d{\dif}

\def\SS{\Sigma}

\def\EE{E}
\def\rr{r}
\def\tt{\theta}

\def\p{P}
\def\x{X}
\def\hp{\hat{P}}
\def\hx{\hat{X}}

\def\af{\zeta}
\def\dr{\delta r}
\def\dt{\delta \theta}

\def\GG{G}
\def\RS{R}

\def\der{|}

\def\mir{\mathrm{r}}
\def\mit{\mathrm{\theta}}

\begin{document}
\title{
String loops oscillating in the field of Kerr black holes as a possible explanation of twin high-frequency quasiperiodic oscillations observed in microquasars
}
\author{Z. Stuchl\'{\i}k}
\author{M. Kolo\v{s}}
\affiliation{Institute of Physics, Faculty of Philosophy \& Science, Silesian University in Opava, Bezru\v{c}ovo n\'{a}m\v{e}st\'{i} 13, CZ-74601 Opava, CzechRepublic}

\begin{abstract}
Small oscillations of current-carrying string loops around stable equilibrium positions corresponding to minima of the effective potential in the equatorial plane of the Kerr black holes are studied using the perturbation method. In the lowest approximation, two uncoupled harmonic oscillators are obtained that govern the radial and vertical oscillations; the higher-order terms determine non-linear phenomena and transition to chaotic motion through quasi-periodic stages of the oscillatory motion. The radial profiles of frequencies of the radial and vertical harmonic oscillations that are relevant also in the quasi-periodic stages of the oscillatory motion are given, and their properties in dependence on the spin of the black holes and the angular momentum and tension of the string loops are determined. It is shown that the radial profiles differ substantially from those corresponding to the radial and vertical frequencies of the geodetical epicyclic motion; however, they have the same mass-scaling and  their magnitude is of the same order. Therefore, we are able to demonstrate that assuming relevance of resonant phenomena of the radial and vertical string-loop oscillations at their frequency ratio $3:2$, the oscillatory frequencies of string loops can be well related to the frequencies of the twin high-frequency quasi-periodic oscillations (HF QPOs) observed in the microquasars GRS 1915+105, XTE 1550-564, GRO 1655-40. We can conclude that oscillating current-carrying string loops have to be considered as one of the possible explanations of the HF QPOs occurring in the field of compact objects. 
\end{abstract}
  
\maketitle

\section{Introduction}\label{intro}

Relativistic current-carrying string loops moving axisymmetrically along the symmetry axis of the Kerr or \Schw\nnd\dS{} black holes have been recently studied extensively \cite{Jac-Sot:2009:PHYSR4:,Kol-Stu:2010:PHYSR4:,Stu-Kol:2012:PHYSR4:,Kol-Stu:2013:PHYSR4:}; the current-carrying string loops were first studied in \citep{Lar:1994:CLAQG:,Fro-Lar:1999:CLAQG:}. Tension of such string loops prevents their expansion beyond some radius, while their worldsheet current introduces an angular momentum barrier preventing them from collapsing into the black hole. There is an important possible astrophysical relevance of the current-carrying string  loops \cite{Jac-Sot:2009:PHYSR4:} as they could in a simplified way represent plasma that exhibits associated string-like behavior via dynamics of the magnetic field lines in the plasma \cite{Chri-Hin:1999:PhRvD:,Sem-Dya-Pun:2004:Sci:} or due to thin isolated flux tubes of magnetized plasma that could be described by an one-dimensional string \cite{Spr:1981:AA:,Sem-Ber:1990:ASS:,Cre-Stu:2013:PhRvE:}. Motion of electrically charged string loops in combined external gravitational and electromagnetic fields has been recently studied for a Schwarzschild black hole immersed in a homogeneous magnetic field \cite{Arm-etal:2013:PRD:}.

The astrophysical applications of the current-carrying string loops have been focused on the problem of acceleration of string loops due to the transmutation process when energy of the oscillatory motion of the string is converted to energy of its translational motion \cite{Jac-Sot:2009:PHYSR4:}. Since the string loops can be accelerated to ultra-relativistic velocities in the deep gravitational potential well of compact objects \cite{Stu-Kol:2012:PHYSR4:,Stu-Kol:2012:JCAP:}, the string loop transmutation can be well considered as a process of formation of ultra-relativistic jets, along with the standard model based on the Blandford-Znajek process \cite{Bla-Zna:1977:MNRAS:} and recently introduced "geodesic collimation" model \cite{Gar-etal:2010:ASTRA:,Pach-etal:2012:ApJ:,Gar-etal:2013:ApJ:}. It has been demonstrated that the cosmic repulsion plays an important role in the acceleration process of the string loops behind the so called static radius of the central object \cite{Stu:1983:BAC:,Stu-Hle:1999:PHYSR4:,Stu-Kol:2012:PHYSR4:}. 

Here we concentrate our attention on the inverse situation of small oscillations of string loops in vicinity of stable equilibrium positions in the equatorial plane of black-hole spacetimes that was proposed as a possible model of HF QPOs observed in black hole and neutron star binary systems \cite{Stu-Kol:2012:JCAP:}. 

In the~black hole systems observed in both Galactic and extragalactic sources, strong gravity effects have a~crucial role in three phenomena related to the~accretion disc that is the~emitting source: the~spectral continuum, spectral profiled lines, and oscillations of the~disc; clearly, strong gravity  has an~important role also in the~binary systems containing neutron (quark) stars. The best signature  of the processes occurring in the strong gravity is frequency of observed oscillations because of possibility to obtain very high precision of its measurements \cite{Fer-etal:2012:ExpAstr:}. HF QPOs of X-ray brightness had been observed in many Galactic Low Mass X-Ray Binaries (LMXB) containing neutron~stars \citep[see, e.g.,][]{Kli:2000:ARASTRA:,Bar-Oli-Mil:2005:MONNR:,Bel-Men-Hom:2007:MONNR:BriNSQPOCor} or black holes \citep[see, e.g.,][]{McCli-Rem:2004:CompactX-Sources:,Rem:2005:ASTRN:,Rem-McCli:2006:ARASTRA:,McCli-Rem:2011:CLAQG:}. Some of the~HF~QPOs are in the~kHz range and often come in pairs of the~upper and lower frequencies ($\nu_{\mathrm{U}}$, $\nu_{\mathrm{L}}$) of {\it twin peaks} in the~Fourier power spectra. 

The~resonance orbital model of HF QPOs in black hole systems \cite{Tor-etal:2005:ASTRA:}, based on the frequencies of the geodetical orbital and epicyclic motion, is now partially supported by observations, in particular when frequency ratio~3\,:\,2 ($2\nu_{\mathrm{U}} = 3\nu_{\mathrm{L}}$) is seen in twin peak QPOs in the~LMXB containing black holes (microquasars), namely GRO 1655-40, XTE 1550-564, GRS 1915+105 \cite{Tor-etal:2005:ASTRA:}. However, in the case of the GRS 1915+105 source the HF QPO frequency set is more complex - in fact, at least five HF QPOs were observed there \cite{McCli-etal:2006:ARAA:}. Therefore, in this case more complex models of HF QPOs have to be considered. In fact, the complete observed frequency set can be explained in the framework of the extended resonant orbital model (\cite{Stu-Sla-Tor:2007a:ASTRA:,Stu-Sla-Tor:2007b:ASTRA:}) based on the so called Aschenbach effect (\cite{Asch:2004:ASTRA:,Stu-Sla-Tor-Abr:2005:PHYSR4:}); another possibility is related to the multi-resonance orbital model recently proposed in \cite{Stu-Kot-Tor:2013:ASTRA:}. 

Nevertheless, there remains a clear problem with explanation of the $3:2$ frequency ratios observed in all three microquasars, GRO 1655-40, XTE 1550-564, GRS 1915+105, if the resonance orbital model with a unique variant of the twin oscillating modes with geodetical frequencies is applied, especially when the limits on the black hole spin given by the spectral continuum fitting (\cite{Rem-McCli:2006:ARASTRA:,McCli-Rem:2011:CLAQG:}) are taken into account \cite{Tor-etal:2011:ASTRA:,Ali-etal:2013:CLAQG:,Ste-Gyu-Yaz:2013:PHYSR4:}. It seems that neither the resonant orbital (geodetical) model or any other proposed model could work simultaneously for all of the three microquasars \cite{Tor-etal:2011:ASTRA:}. 

Therefore, we will test, whether the frequencies of the $3:2$ twin peak oscillations observed in the three microquasars can be explained by the axisymmetric current-carrying string loops oscillating in the field of a Kerr black hole, if the oscillations occur at the "resonant" radii where the radial and vertical frequencies have the rational ratio $3:2$. In Section 2 we summarize the Hamiltonian formalism for the string loop motion and give the perturbative form of the Hamiltonian for motion in vicinity of equilibrium points located in the equatorial plane of Kerr black holes. In Section 3 we give the radial profiles of the frequencies of the radial and vertical modes of the oscillatory motion in terms of the dimensionless spin of the black hole and the angular momentum parameter of the string loop. We discuss their properties, comparing them to the radial profiles of the geodesic radial and vertical epicyclic motion in the Kerr backgrounds. In Section 4 we apply the oscillatory string-loop model to explain the  frequencies of the twin peak HF QPOs observed with the frequency ration $3:2$ in the three microquasars. Concluding remarks are presented in Section 5. 

\section{Dynamics of relativistic current-carrying string loops in the field of Kerr black holes}

The equations of motion for relativistic current-carrying string loop with tension $\mu$ and scalar field $\phi$ governing the current generating an angular momentum are in the standard form presented in \cite{Jac-Sot:2009:PHYSR4:,Kol-Stu:2013:PHYSR4:}. The Hamiltonian form is introduced by \cite{Lar:1994:CLAQG:} and for the Kerr spacetimes it is discussed in detail in the work \cite{Kol-Stu:2013:PHYSR4:}, the results of which are summarized in the following. 

\subsection{Kerr geometry}
Kerr black holes are described by the Kerr geometry that is given in the standard Boyer-Lindquist coordinates and the geometric units ($c=G=1$) in the form
\bea
 \d s^2 &=& - \left( 1- \frac{2Mr}{\RS^2} \right) \d t^2 - \frac{4Mra \sin^2\theta}{\RS^2} \, \d t \d \phi \nonumber\\
 && + \left( r^2 +a^2 + \frac{2Mra^2}{\RS^2} \sin^2\theta \right) \sin^2\theta \, \d \phi^2 \nonumber \\
 && + \frac{\RS^2}{\Delta} \, \d r^2 + \RS^2\, \d\theta^2, 
 \label{KerrMetric} 
\eea
where
\beq
\RS^2 = r^2 + a^2 \cos^2\theta, \quad \Delta = r^2 - 2Mr + a^2,
\eeq
$a$ denotes spin and $M$ mass of the spacetimes. The physical singularity is located at the ring $r=0, \theta = \pi/2$. For $a>M$, the Kerr geometry describes naked singularity spacetimes, for $a<M$, black hole spacetimes. In the present paper we consider only the Kerr black hole spacetimes at the external region located above the event horizon ($r > r_{+}$) where the ring singularity and the causality violating region are irrelevant. The outer horizon is given by  
\beq
                 r_{+} = M + (M^2-a^2)^{1/2}.
\eeq
The stationary limit surface $r_{\rm stat}(\theta)$, governing the boundary of the ergosphere, is given by 
\beq
              r_{\rm stat}(\theta) = M + (M^2-a^2 \cos^2\theta)^{1/2} .
\eeq

We shall use only the Boyer-Lindquist coordinates, since the Kerr-Schild "Cartesian" coordinates are useful in close vicinity of the ring singularity \cite{Kol-Stu:2013:PHYSR4:}. 
%
%
Using dimensionless radial coordinate $r\rightarrow r/M$, dimensionless time coordinate $t\rightarrow t/M$ and dimensionless spin $a\rightarrow a/M$, we can exclude mass $M$ from our equations. We will return to the dimensional quantities in the section \ref{observations}.
%
%

\subsection{Hamiltonian formulation of the equations of motion of string loops}

The string worldsheet is described by the spacetime coordinates $X^{\alpha}(\sigma^{a})$ with $\alpha = 0,1,2,3$ given as functions of two worldsheet coordinates $\sigma^{a}$ with $a = 0,1$ that imply induced metric on the worldsheet in the form
\beq
      h_{ab}= g_{\alpha\beta}X^\alpha_{\der a}X^\beta_{\der b},
\eeq 
where $\Box_{\der a} = \partial \Box /\partial a$. The string current localized on the worldsheet is described by a scalar field $\phi({\sigma^a})$. Dynamics of the string, inspired by an effective description of superconducting strings representing topological defects occurring in the theory with multiple scalar fields undergoing spontaneous symmetry breaking \cite{Wit:1985:NuclPhysB:,Vil-She:1994:CSTD:}, is described by the action $S$ with Lagrangian $\mathcal{L}$
\beq
 S = \int \mathcal{L} \, \dif \sigma \dif \tau, \quad 
 \mathcal{L} = -(\mu + h^{ab} \varphi_{\der a}\varphi_{\der b})\sqrt{-h}, \label{katolicka_akce}
\eeq
where $ \varphi_{,a} = j_a $ determines current of the string and $\mu > 0$ reflects the string tension.

Varying the action with respect to the induced metric $h_{ab}$ yields the worldsheet stress-energy tensor density (being of density weight one with respect to worldsheet coordinate transformations)
\beq
\SS^{ab}= \sqrt{-h} \left( 2 j^a j^b - (\mu + j^2) h^{ab}\right),
\eeq
where
\beq
   j^a = h^{ab}j_{b} , \quad j^2 = h^{ab}j_{a}j_{b}.
\eeq
The contribution from the string tension with $\mu > 0$ has a positive energy density and a negative pressure (tension). The current contribution is traceless, due to the conformal invariance of the action - it can be considered as a $1+1$ dimensional massless radiation fluid with positive energy density and equal pressure \cite{Jac-Sot:2009:PHYSR4:}. 

In the conformal gauge, the equation of motion of the scalar field reads (see \cite{Jac-Sot:2009:PHYSR4:} for details)
\beq
    \varphi_{\der \tau \tau} - \varphi_{\der \sigma \sigma} = 0.
\eeq
The assumption of axisymmetry implies that the current is independent of $\sigma$ and $j_{a,\sigma} = 0$. Using the scalar field equation of motion we can conclude that the scalar field can be expressed in a linear form with constants $j_{\sigma}$ and $j_{\tau}$
\beq
    \varphi = j_{\sigma}\sigma + j_{\tau}\tau.
\eeq
Introducing new quantities 
\beq
     J^2 \equiv j_\sigma^2 + j_\tau^2, \quad \omega \equiv -j_\sigma / j_\tau ,
\eeq
we express the components of the worldsheet stress-energy density $\SS^{ab}$ in the form
\bea
&& \SS^{\tau\tau} = \frac{J^2}{g_{\phi\phi}} + \mu , \quad \SS^{\sigma\sigma} = \frac{J^2}{g_{\phi\phi}} - \mu , \label{sigma1}\\ 
&& \SS^{\sigma\tau} =  \frac{-2 j_\tau j_\sigma}{g_{\phi\phi}} = \frac{2 \omega J^2}{g_{\phi\phi} (1+\omega^2)}. \label{sigma2}
\eea
The string dynamics depends on the current through the worldsheet stress-energy tensor. The dependence is expressed using the angular momentum parameters $J^2$ and $\omega$. The minus sign in the definition of $\omega$ is chosen in order to obtain correspondence of positive angular momentum and positive $\omega$. For a given string loop, the tension $\mu>0$ and both angular momentum parameters, $J>0$ and $\omega\in\langle-1,1\rangle$, are constant during its motion. 

In the Kerr spacetime we can introduce the locally non-rotating frames corresponding to zero-angular-momentum observers (ZAMO) that are co-moving with the spacetime rotation \cite{Bar:1973:BlaHol:}. ZAMO are observing the string loop in coordinates
\beq
 X^\alpha(\tau,\sigma) = (t(\tau),r(\tau),\theta(\tau),\sigma + f(\tau)). \label{strcoord}
\eeq
Now it is clear that relative to ZAMO the string loops do not rotate, and for the string coordinates (\ref{strcoord}) we can obtain the relations 
\bea
 \dot{X}^\alpha &=& X^\alpha_{\der\tau} = (t_{\der\tau}\,,r_{\der\tau}\,,\theta_{\der\tau}\,,f_{\der\tau}\,), \\
 {X'}^\alpha &=& X^\alpha_{\der\sigma} = (0,0,0,1).
\eea
For the function $f(\tau)$, we obtain
\beq
 f_{\der\tau} = -(g_{t\phi}/g_{\phi\phi}) t_{\der\tau}. 
\eeq

\begin{figure}
\includegraphics[width=0.8\hsize]{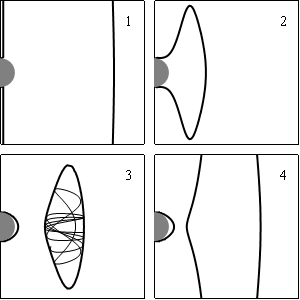}
\caption{\label{string_clasBH} Four different types of the behavior of the boundary energy function $E_{\rm b}(x,y;J)$ in the Kerr black hole spacetimes; for details see \cite{Jac-Sot:2009:PHYSR4:, Kol-Stu:2013:PHYSR4:}. 
}
\end{figure}

Varying the action with respect to $X^\mu$ implies equations of motion in the form
\beq
 \frac{\rm D}{\d \tau} P^{(\tau)}_\mu + \frac{\rm D}{\d \sigma} P^{(\sigma)}_\mu = 0, \label{EQmotion}
\eeq
where the string loop momenta are defined by the relations
\bea
 P^{(\tau)}_\mu &\equiv& \frac{\partial \mathcal{L}}{\partial \dot{X}^\mu} = \SS^{\tau a} g_{\mu \lambda} X^\lambda_{\der a}, \\
 P^{(\sigma)}_\mu &\equiv& \frac{\partial \mathcal{L}}{\partial {X'}^\mu} = \SS^{\sigma a} g_{\mu \lambda} X^\lambda_{\der a}.
\eea

Defining affine parameter $\af$, related to the worldsheet coordinate $\tau$ by the transformation  
\beq
 \d \tau = {\SS^{\tau\tau}} \d \af ,
\eeq
we can find the Hamilton equations
\beq
 \frac{\d \x^\mu}{\d \af} = \frac{\partial H}{\partial \p_\mu}, \quad
 \frac{\d \p_\mu}{\d \af} = - \frac{\partial H}{\partial \x^\mu}, \label{Ham_eq}
\eeq
for the Hamiltonian given in the form
\beq
 H = \frac{1}{2} g^{\alpha\beta} \p_\alpha \p_\beta + \frac{1}{2} g_{\phi\phi} \left[(\SS^{\tau\tau})^2 - (\SS^{\tau\sigma})^2 \right] \label{AllHam}
\eeq
where $\alpha, \beta$ are the spacetime coordinates $ t,r,\theta,\phi$ -- see \cite{Kol-Stu:2013:PHYSR4:} for details. 

\begin{figure*}
\includegraphics[width=\hsize]{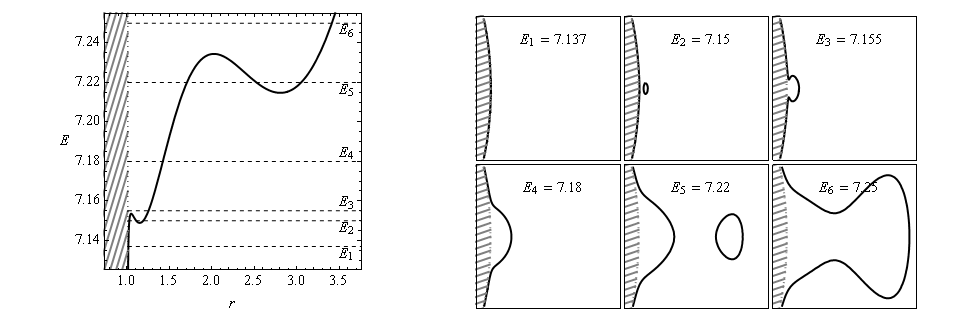}
\caption{\label{fceEB}
Energy boundary function $E_{\rm b}(x,y;a,J,\omega)$ constructed for the representative case of the near-extreme black holes, regions below the horizon are hatched. Both the radial profile and the meridional sections for representative values of constant energy are given for $a=0.9999, \omega=0.35, J=4.8$. }
\end{figure*}

\subsection{Integrals of the motion and the energy boundary function}

The Kerr metric (\ref{KerrMetric}) does not depend on coordinates $t$ (stacionarity) and $\phi$ (axial symmetry). Such symmetries  imply conserved quantities - string energy and axial angular momentum. The conserved string energy $E$ takes the form 
\beq
 - E = P_t = \SS^{\tau\tau} \left(g_{tt} -g_{t\phi}^2/g_{\phi\phi} \right) t_{\der\tau} + g_{t\phi} \SS^{\sigma\tau}. \label{Cenergy}
\eeq
The string loops do not rotate, but they have non-zero angular momentum, generated by the current. The axial component of the angular momentum reads
\beq
 L = P_\phi = g_{\phi\phi} \SS^{\sigma\tau} = -2 j_\tau j_\sigma. \label{Cmoment}
\eeq
From equations (\ref{sigma1}-\ref{sigma2}) we clearly see that $\omega<0$ represents string loops with negative axial angular momentum $L<0$, $\omega=~0$ represents the string loops with $L=0$, and $\omega>0$ represents the string loops with $L>0$. Recall that in the non-rotating Schwarzschild spacetime, or any spherically symmetric spacetime, the string loop equations of motion are degenerated, being independent of the parameter $L$ (or $\omega$) \cite{Kol-Stu:2010:PHYSR4:,Stu-Kol:2012:JCAP:}. 

The string dynamics depends on the current $J$ through the worldsheet stress-energy tensor. Using the two constants of motion (\ref{Cenergy}-\ref{Cmoment}), we can rewrite the Hamiltonian (\ref{AllHam}) into the form related to the $r$ and $\theta$ momentum components
\bea
 H &=& \frac{1}{2} g^{rr} \p_r^2 + \frac{1}{2} g^{\theta\theta} \p_\theta^2 \nonumber\\
 && + \frac{1}{2} g_{\phi\phi} (\SS^{\tau\tau})^2 + \frac{g_{\phi\phi}(E + g_{t\phi} \SS^{\sigma\tau})^2}{2 (g_{tt} g_{\phi\phi}-g_{t\phi}^2)}. \label{HamHam}
\eea
The equations of motion (\ref{Ham_eq}) following from the Hamiltonian (\ref{HamHam}) are very complicated and can be solved only numerically in general case, although there exist analytical solutions for simple cases of the motion in the flat or de~Sitter spacetimes \cite{Kol-Stu:2010:PHYSR4:}.

The loci where a string loop has zero velocity ($\dot{r}=0, \dot{\theta}=0$) form boundary of the string motion that can be appropriately related to its energy. The boundary energy function serves as an effective potential of the string loop motion and is defined by the relation 
\beq
 E = E_{\rm{b}}(r,\theta) = \sqrt{g_{t\phi}^2-g_{tt}g_{\phi\phi}} \, \SS^{\tau\tau} - g_{t\phi}\SS^{\sigma\tau} \label{EqEbRT} \label{StringEnergy},
\eeq  
where we use condition $H=0$, given by reparameterization invariance of the action (\ref{katolicka_akce}) \cite{Kol-Stu:2013:PHYSR4:}.

We can make the rescaling $E_{\rm b} \rightarrow E_{\rm b} / \mu $ and $J \rightarrow J / \sqrt{\mu} $,  due to the assumption of $\mu > 0$. This choice of ``units'' will not affect energy boundary function and is equivalent to setting the string tension $\mu=1$ in Eqs (\ref{sigma1}-\ref{sigma2}), (\ref{EqEbRT}). The energy boundary function then, in the standard Boyer-Lindquist $r,\theta$ coordinates, and in terms of the angular momentum parameters $J$ and $\omega$, takes the form  
\beq
 E_{\rm b} (r,\theta;a,J,\omega) = \frac{4 a \omega J^2 r }{\left(\omega^2+1\right) \GG }+\sqrt{\Delta}
   \left(\frac{J^2 R^2}{\GG \sin(\theta)}+\sin(\theta)\right), \label{EBrtKerr}
\eeq
where we introduced the function
\beq    
  \GG (r,\theta;a) = \left(a^2+r^2\right) R^2 +2 a^2 r \sin^2(\theta ). \label{Gfce}
\eeq

We define the function $J^2_{\rm E}(r;a,\omega)$ governing the stationary points (local extrema) of the energy boundary function $E_{\rm b}(r;a,J,\omega)$ in the equatorial plane ($\theta=\pi/2$) by the relation \cite{Kol-Stu:2013:PHYSR4:}
\beq
J^2_{\rm E}(r;a,\omega) =  \frac{(r-1) \left(\omega^2+1\right) H^2}{4 a \omega \sqrt{\Delta} \left(a^2+3 r^2\right)+\left(\omega^2+1\right) F }, \label{eqrovina}
\eeq
where we introduced the functions
\bea
 H(r;a) &=& r^3 + a^2 (2 + r),\\
 F(r;a) &=& (r-3) r^4 -2 a^4+a^2 r \left(r^2-3 r+6\right). \label{Ffce}
\eea
We restrict our investigation to the black hole spacetimes ($0 \leq a \leq 1$) and the regions located above the outer horizon. The local extrema of the $J^2_{\rm E}(r;a,\omega)$ function, given by the condition $(J^2_{\rm E}\,)_{r}'~=~0$, enable us to distinguish the maxima and minima of the energy boundary function $E_{\rm b}(r;a,J,\omega)$. The local extrema of the function $J^2_{\rm E}(r;a,\omega)$ are given by the relation
\beq
 J_{\rm E(ex)}(r,a,\omega) = 0 \label{JEex0}
\eeq
where 
\begin{widetext}
\bea
J_{\rm E(ex)}(r;a,\omega) &\equiv& \left(\omega ^2+1\right) \left[ H (r-1) \left( 6 a^2 r -3 a^2 r^2 -6 a^2 -5 r^4+ 12 r^3 \right)- F H -2 F (a^2+3r^2) (r-1)\right] \nn \\
&& +4 a \omega \Delta^{-1/2} \left[H (a^2+3r^2) \left(\Delta -(r-1)^2\right)-6 \Delta r H (r-1) +2 \Delta  (a^2+3r^2)^2 (r-1)\right]. \label{JEex}
\eea 
\end{widetext}
We can see that the function $J_{\rm E(ex)}(r;a,\omega)$ is a very complex high-order polynomial function of the radial coordinate $r$ and the black hole spin $a$, but it is only quadratic function of the parameter $\omega$. 

There are four different types of the behavior of the energy boundary function for the string loop dynamics in the Kerr BH or \Schw{} spacetimes represented by the characteristic $E={\rm{}const}.$ sections of the function $E_{\rm b}(r,\theta)$ depending on the parameters $J,\omega,a$ \cite{Jac-Sot:2009:PHYSR4:}. We can distinguish them according to two properties: possibility of the string loop to escape to infinity in the $y$-direction, and possibility to collapse to the black hole. A detailed discussion can be found in \cite{Kol-Stu:2013:PHYSR4:}; here we shortly summarize the results. 

In the Kerr BH spacetimes, we can distinguish four different types of the behavior of the boundary energy function and the character of the string loop motion; in Fig. \ref{string_clasBH} we denote them by points numbered by 1 to 4. The first case corresponds to no inner and outer boundary - the string loop can be captured by the black hole or escape to infinity. The second case corresponds to the situation with an outer boundary - the string loop must be captured by the black hole. The third case corresponds to the situation when both inner and outer boundary exist - the string loop is trapped in some region forming a potential ``lake'' around the black hole. The fourth case corresponds to an inner boundary - the string loop cannot fall into the black hole but it must escape to infinity. 

\begin{figure*}
\includegraphics[width=\hsize]{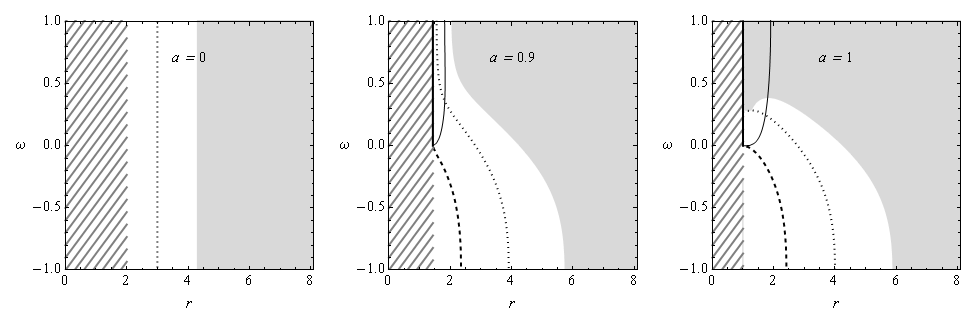}
\caption{\label{TTsitOM}
Properties of the fundamental angular frequency of the vertical harmonic oscillations $\Omega_{\mit}(r;a,\omega)$ determined by the radial profiles of the characteristic functions $\omega(r;a)$ that are given for three typical values of the black hole spin $a\in\{0,0.9,1\}$. Regions where the string-loop motion is bounded and the expressions for the harmonic oscillations are relevant are greyed, regions below the horizon are hatched. 
The dashed curves give the function $\omega_{\rm d}(r;a)$ determining position of divergence of the  $\Omega_{\mit}(r;a,\omega)$ function, the dotted curves give the allowance function $\omega_{\rm a}(r;a)$ determining region where the condition $J^2_{\rm E}(r)\geq~0$ is satisfied, the thin curves give the  extrema function $\omega_{\rm e(\mit)}(r;a)$ determining loci where $\d \Omega_{\mit} / \d r = 0$.
The zero points of the $\Omega_{\mit}(r;a,\omega)$ function are for the black holes (with $a<1$)
and $\omega>0$ formally located at the outer horizon.
}
\end{figure*}

For our following discussion only the third case, corresponding to the existence of "lakes" in the effective potential (energy boundary function) given for fixed values of the angular momentum parameters $J$ and $\omega$, will be relevant. The lakes are governed by the $E={\rm{}const}.$ sections of the energy boundary function, at their center a stable equilibrium position of a string loop with fixed angular momentum parameters occurs, having energy corresponding to the minimal allowed value. Around such equilibrium positions, small oscillations of the string loop occur if its energy slightly exceeds the minimal value. 


\subsection{Perturbative Hamiltonian near equilibrium points}

The equilibrium points of Hamiltonian (\ref{HamHam}) correspond to the local minima at $\x^\alpha_0=(\rr_0,\tt_0)$ of the energy boundary function $\EE_{\rm b}(\rr,\tt)$. Therefore, the Kolmogorov-Arnold-Moser theorem \citep{Arnold:1978:book:} can be applied. It is useful to write the  Hamiltonian in the form
\beq
 H = H_D + H_P = \frac{1}{2} g^{rr} \p_r^2 + \frac{1}{2} g^{\theta\theta} \p_\theta^2 + H_P(r,\theta)
\eeq 
where we split $H$ into "dynamical" $H_{\rm D}$ and "potential" $H_{\rm P}$ parts. Introducing a small parameter $\epsilon << 1$, we can rescale coordinates and momenta by the relations 
\beq
 \x^\alpha = \x^\alpha_0 + \epsilon \hx^\alpha, \quad \p_\alpha = \epsilon \hp_\alpha,
\eeq
applied for the coordinates $\alpha \in \{ \rr,\tt \}$. The polynomial expansion of the Hamiltonian into Taylor series around an equilibrium point and expressing in powers of the small parameter $\epsilon$ gives 
\bea
 H (\hp_\alpha,\hx^\alpha) &=& H_0 + \epsilon H_1(\hx^\alpha) + \epsilon^2 H_2(\hp_\alpha,\hx^\alpha) \nonumber\\
  && + \epsilon^3 H_3(\hp_\alpha,\hx^\alpha) + \ldots, \label{expand1} 
\eea
where $H_k$ is homogeneous part of the Hamiltonian of degree $k$ considered for the momenta $\hp_\alpha$ and coordinates $\hx^\alpha$. Recall that $\p_\alpha$ is quadratic in (\ref{HamHam}) and apears in $H_k$ only for $k \geq 2$. If the string is sitting in local minima of $\EE_{\rm b}$, we have $H_{\rm D} = 0$ and hence $H_0 = 0$. The local extrema of the energy boundary function $\EE_{\rm b}$ imply also $H_1(\hx^\alpha) = 0$.

We can divide (\ref{expand1}) by the factor $\epsilon^2$ (recall that $H=0$) expressing the Hamiltonian in the vicinity of the local minimum in the "regular" plus "perturbation" form 
\beq
 H =  H_2(\hp_\alpha,\hx^\alpha) + \epsilon H_3(\hp_\alpha,\hx^\alpha) + \ldots
\eeq
If the parameter $\epsilon = 0$, we arrive to an integrable Hamiltonian
\beq
 H = H_2(\hp_\alpha,\hx^\alpha) = \frac{1}{2} \sum_\alpha \left[ g_{\alpha\alpha}(\hp_\alpha)^2 + \omega^2_\alpha (\hx^\alpha)^2 \right]
\eeq
representing two uncoupled linear harmonic oscillators \cite{Kol-Stu:2013:PHYSR4:}. Performed "perturbation" approach corresponds to the linearization of motion equations in the neighborhood of the  minima of the energy boundary function $E_{\rm b}(r;a,J,\omega)$. 

\begin{figure*}
\includegraphics[width=\hsize]{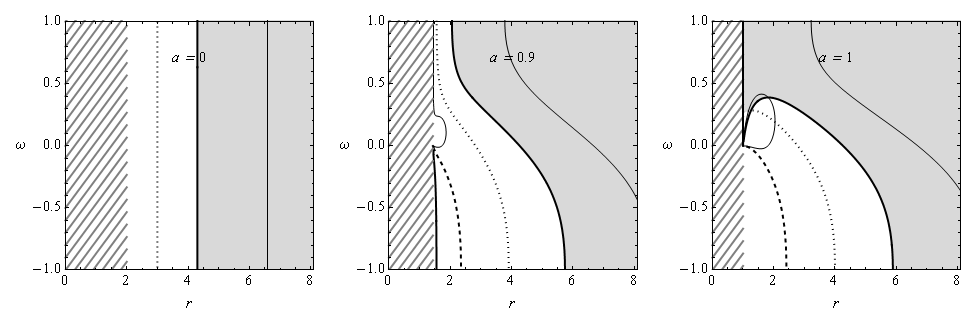}
\caption{\label{RRsitOM}
Properties of the fundamental angular frequency of the radial harmonic oscillations $\Omega_{\mir}(r;a,\omega)$ determined by the radial profiles of the characteristic functions $\omega(r;a)$ that are given for three typical values of the black hole spin $a\in\{0,0.9,1\}$. Regions where the string-loop motion is bounded and the expressions for the harmonic oscillations are relevant are greyed, regions below the outer horizon are hatched. 
The dashed curves give the divergence function $\omega_{\rm d}(r;a)$, the dotted curves give the allowance function $\omega_{\rm a}(r;a)$, the thin curves give the extrema function $\omega_{\rm e(\mir)}(r;a)$ determining loci where $\d \Omega_{\mir} / \d r = 0$. 
Thick black curves give the zero function $\omega_{\rm z(\mir)}(r;a)$ determined by the condition $\Omega_{\rm r}=0$ that determines the region of the bounded energy constant sections allowing for both the radial and vertical oscillations.
}
\end{figure*}

\section{Frequencies of harmonic oscillations of string loops}

Assuming small oscillations of an axisymmetric string loop near a stable equilibrium position  corresponding to a minimum of the effective potential, the Hamiltonian of its motion has been given in a perturbative form, with leading term corresponding to linear harmonic oscillators in two uncoupled orthogonal modes - the radial and vertical ones. The higher order terms, characterized by the small parameter $\epsilon$, govern non-linear phenomena causing coupling of the radial and vertical oscillatory modes, and determine transition to chaotic motion through quasi-periodic stages of the oscillatory motion. The frequencies of the radial and vertical harmonic oscillations are relevant also in the quasi-periodic stages of the oscillatory motion \cite{Nay-Moo:1979:NonOscilations:,Stu-Kol:2012:JCAP:}. Now we give the frequencies of the harmonic oscillations at the equatorial plane of Kerr black holes and determine properties of their radial profiles. 

For harmonic oscillations of string loops around a stable equilibrium position with fixed coordinates $r_0$ and $\theta_0=\pi/2$ (recall that in the Kerr black hole spacetimes the equilibrium positions are allowed at the equatorial plane only) the string loop coordinates $\rr = \rr_0 + \dr,\tt = \tt_0 + \dt$ are governed by the equations  
\beq
 \ddot{\dr} + \omega^2_{\mir} \, \dr = 0, \quad \ddot{\dt} + \omega^2_{\mit} \, \dt = 0,
\eeq
where the locally measured angular frequencies of the oscillatory motion are given by 
\beq
 \omega^2_{\mir}  =  \frac{1}{g_{rr}} \, \frac{\partial^2 H_P}{\partial r^2}, \quad 
 \omega^2_{\mit}  =  \frac{1}{g_{\tt\tt}} \, \frac{\partial^2 H_P}{\partial \theta^2}.
\eeq

The locally measured angular frequencies 
\beq
\omega_{(\mir,\mit)}= \frac{\d f_{(\mir,\mit)}}{\d \af}
\eeq
are connected to the angular frequencies related to distant observers, $\Omega$, by the gravitational redshift transformation
\beq
 \Omega_{(\mir,\mit)} = \frac{\d f_{ (\mir,\mit)}}{\d t} = \frac{\omega_{(\mir,\mit)}}{P^t} ,
\eeq
where
\beq
 P^t = \frac{\d t}{\d \af} = - g^{tt} E + g^{\phi t} L.
\eeq
If the angular frequencies $\Omega_{(\mir,\mit)}$, or frequencies $\nu_{(\mir,\mit)}$, of the string loop oscillation are expressed in the physical units, their dimensionless form has to be extended by the factor $c^3/GM$. Then the frequencies of the string loop oscillations measured by the distant observers are given by 
\beq
     \nu_{(\mir,\mit)} = \frac{1}{2\pi} \frac{c^3}{GM} \, \Omega_{(r,\theta)}.
\eeq
Notice that this is the same factor as the one occurring in the case of the orbital and epicyclic frequencies of the geodetical motion in the black hole spacetimes \cite{Tor-Stu:2005:ASTRA:}. Therefore, the order of magnitude and scaling of the frequencies of the radial and vertical oscillations due to the mass of the central object is the same for both current-carrying string loops and test particles. 

The angular frequencies of the string loop oscillations related to distant observers take the dimensionless form 
\begin{widetext}
\bea
 \Omega^2_{\mir}(r;a,\omega) &=& \frac{ J_{\rm E(ex)} \, \left(2 a \omega \sqrt{\Delta } \left(a^2+3 r^2\right)+\left(\omega ^2+1\right) \left(a^2 r^3 -a^2 \Delta  +r^5-2 r^4\right)\right)}{2 r \left(a^2
   (r+2)+r^3\right)^2 \left(2a\omega  \left(a^2+3r^2\right)+\sqrt{\Delta } \left(\omega ^2+1\right) \left(r^3-a^2\right)\right)^2} , \\
  \Omega^2_{\mit}(r;a,\omega) &=& \frac{\sqrt{\Delta } \left(2 a \omega \sqrt{\Delta }   \left(2 a^2 -3 a^2 r -3 r^3\right)+\left(\omega ^2+1\right) \left(a^4 (3 r-2)+2 a^2 (2 r-3) r^2+r^5\right)\right)}{r^2
   \left(a^2 (r+2)+r^3\right) \left(2a\omega  \left(a^2+3r^2\right)+\sqrt{\Delta } \left(\omega ^2+1\right) \left(r^3-a^2\right)\right)} . 
\eea
\end{widetext}
For $\omega=0$, these very complex relations take a simpler form 
\bea
\Omega^2_{\mir}(r;a) &=& \frac{ J_{\rm E(ex)} \, \left(a^2 (r^3-\Delta) +r^5-2 r^4\right)}{2 \Delta  r \left(a^2-r^3\right)^2 \left(a^2 (r+2)+r^3\right)^2}, \\
\Omega^2_{\mit}(r;a) &=& \frac{2 a^4 -3a^4 r -4 a^2 r^3+6 a^2 r^2-r^5}{r^2 \left(a^2-r^3\right) \left(a^2 (r+2)+r^3\right)}.
\eea
Even in this simplest case, the relations for the angular frequencies of the harmonic oscillations are still much more complex in comparison with the relations for the angular frequencies of the epicyclic geodetical motion in the field of Kerr black holes that take the form \cite{Tor-Stu:2005:ASTRA:}
\bea
 \Omega^2_{\rm \mir(geo)}(r;a) &=& \frac{-3 a^2+8 a \sqrt{r}+(r-6) r}{r^2 \left(a+r^{3/2}\right)^2} , \\
 \Omega^2_{\rm \mit(geo)}(r;a) &=& \frac{3 a^2-4 a \sqrt{r}+r^2}{r^2 \left(a+r^{3/2}\right)^2} .
\eea
The radial and latitudinal frequencies of the string loop oscillations and the geodetical epicyclic motion are different, enabling some substantial modifications in the relation of the frequency ratio of the twin HF QPOs modeled by the string loop oscillations or the geodetical epicyclic oscillations.

\begin{figure*}
\includegraphics[width=\hsize]{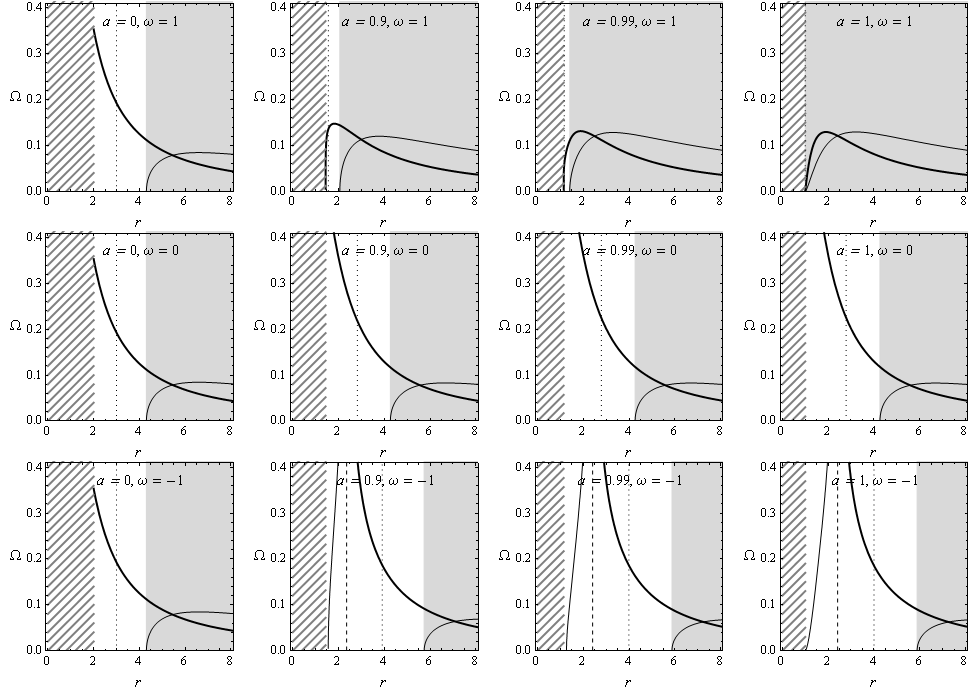}
\caption{\label{exFF}
The radial profiles of the string-loop fundamental angular frequencies $\Omega_\mir$ (thin) and $\Omega_\mit$ (thick). The profiles are given for representative values of the black hole spin $a\in\{0,0.9,0.99,1\}$ and the string-loop parameter $\omega\in\{-1,0,1\}$. The regions where the "lakes" of the constant energy sections of the energy boundary function $E_{\rm b}(r,\theta)$ can exist and the string-loop motion is bounded (being of the type 3 in Fig. \ref{string_clasBH}) is grayed. The regions below the BH outer horizont is hatched, and the divergence point of the angular frequency functions (for $\omega<0$) is denoted by the dashed line. The dotted vertical lines give the "allowance" limit determined by $\omega_{\rm a}(r;a)$. Physically relevant are only the shaded regions where both the radial and vertical angular frequencies are well defined simultaneously.
}
\end{figure*}

In the Schwarzschild spacetimes with $a=0$, the harmonic oscillations have frequencies relative to distant observers given by expressions relatively very simple for both string loops and test particles. Therefore, we can give the frequencies in dimensional form, as an example.
In the case of string loops they read 
\beq
\Omega^2_{\mir}(r) = \frac{3 M^2-5 M r+r^2}{r^4},\quad \Omega^2_{\mit}(r) = \frac{M}{r^3}, \label{OMs} 
\eeq 
while for the epicyclic motion of test particles there is 
\beq
 \Omega^2_{\rm \mir (geo)}(r) = \frac{M (r-6 M)}{r^4}, \quad \Omega^2_{\rm \mit (geo)}(r) = \frac{M}{r^3}. \label{OMp}
\eeq
It is quite interesting that the latitudinal frequency of the string loop oscillations in the \Schw{}  ($a=0$) or other spherically symmetric spacetimes equals to the latitudinal frequency of the epicyclic geodetical motion as observed by distant observers - for details see \cite{Stu-Kol:2012:JCAP:}. Therefore, in the degenerate situation of the spherically symmetric spacetimes, where the parameter $\omega$ is irrelevant for the string loop oscillatory motion, the vertical oscillations are fully governed by the gravity effect of the black hole and the string tension plays no role. However, it is not so in the rotating Kerr spacetimes where, even for string loops with $\omega = 0$, the vertical harmonic oscillation have different form for the string loops and test particles, indicating an important role of the string tension and angular momentum even in the simplest state.

\subsection{Properties of the radial profiles of the angular frequencies of the string-loop harmonic oscillations}

We discuss properties of the radial profiles of both the radial and vertical angular frequencies of the  harmonic oscillations. 

First, we note that the radial profiles of the angular frequencies fulfill the symmetry relations 
\beq
      \Omega_{(\mir,\mit)}(r;a,\omega) = \Omega_{(\mir,\mit)}(r;-a,-\omega) ;  
\eeq
therefore, it is enough to consider the spin parameter $a$ in the interval $0 \leq a \leq 1$, for the axial angular momentum parameter of the string loop in the interval $-1 \leq \omega \leq 1$. For $\omega=0$, the angular frequencies depend on the $a^2$ only, i.e., there is 
\beq
                \Omega_{(\mir,\mit)}(r;a) = \Omega_{(\mir,\mit)}(r;-a) . 
\eeq
Notice that for angular frequencies of the epicyclic geodetical motion $\Omega_{\rm (\mir,\mit)(geo)}(r;a)$ such a symmetry does not hold.

Second, we have to look for the zero points of the radial profiles that determine the limits on the existence of the oscillatory string-loop equilibrium points, and the local extrema of the radial profiles determining extremal values of the observed frequencies. The divergent points of the radial profiles have to be studied too. Moreover, there is also a new feature of the radial profiles of the radial and vertical oscillations of the string loops that is not present in the case of the epicyclic test particle motion and deserves a particular attention, namely the possibility that the radial and vertical frequencies coincide in a particular radius. 

We have to stress that the string-loop harmonic oscillations are possible only if the equilibrium positions of the string loops are allowed, i.e., when the "lakes" of the energy boundary function exist. Therefore, the conditions 
\beq
  J_{\rm E(ex)}=0 \quad \textrm{and} \quad J_{\rm E}^2 \geq 0
\eeq   
have to be satisfied simultaneously, and they put the limit on the validity of the formulas giving the angular frequencies of the radial and vertical oscillations. The second condition for $J_{\rm E}^2$ means that only the positive branch of the solution for $J_{\rm E}^2(r)$ function is relevant \cite{Kol-Stu:2013:PHYSR4:} and implies that the equilibrium points of the string loops will be allowed if 
\beq
        \omega > \omega_{\rm a}(r;a)
\eeq 
where the "allowance" function $\omega_{\rm a}(r;a)$ is a proper solution of the equation $J_{\rm E}^2\geq~0$ implying the condition 
\beq
      4 a \omega \sqrt{\Delta} \left(a^2+3 r^2\right)+\left(\omega^2+1\right) F = 0
\eeq
that is determined by vanishing of denominator of the expression giving $J_{\rm E}^2(r;a,\omega)$, since its nominator is positive above the outer horizon (at $r>1$). 

Now we can study properties of the radial profiles of the angular frequencies $\Omega_{\mit}^2(r;a,\omega)$ and $\Omega_{\mir}^2(r;a,\omega)$ of the string loop harmonic oscillations, taking into account the limits given by the function $\omega_{\rm a}(r;a)$. 

The divergent points of both the functions $\Omega_{\mit}^2(r;a,\omega)$ and $\Omega_{\mir}^2(r;a,\omega)$ are determined by the equation 
\beq
       \omega = \omega_{\rm d}(r;a)
\eeq
where the "divergence" function $\omega_{\rm d}(r;a)$ is given by a proper solution of the equation  
\beq
  2a\omega  \left(a^2+3r^2\right)+\sqrt{\Delta } \left(\omega ^2+1\right) \left(r^3-a^2\right)=0. 
\eeq

The zero points of the frequency profile of the vertical oscillations are given by the condition 
\beq
       \Omega_{\mit}(r;a,\omega) = 0 .
\eeq
Therefore, the zero points can occur at the horizon, or if equality in the condition 
\bea
  && 2 a \sqrt{\Delta } \omega  \left(-3 a^2 r+2 a^2-3 r^3\right) \label{nn} \\  
	&&+\left(\omega ^2+1\right)\left(a^4 (3 r-2)+2 a^2 (2 r-3) r^2+r^5\right) \geq 0 \nn 
\eea
is satisfied. We have found no solution of the equality in the condition (\ref{nn}) that is always positive. The horizon is reached, if a maximum of the function $\Omega_{\mit}(r;a,\omega)$ occurs. However, the function $\Omega_{\mit}(r;a,\omega)$ has to be related to the allowance function 
$\omega_{\rm a}(r;a)$. 

The zero points of the radial profile of the frequency of the radial oscillations are given by the condition 
\beq
       \Omega_{\mir}^2 (r;a,\omega) = 0 
\eeq
that corresponds to the limit on the existence of the stable equilibrium positions of the string loops in the form of the condition $J_{\rm E(ex)}(r;a,\omega)=0$. 
\begin{figure*}
\includegraphics[width=\hsize]{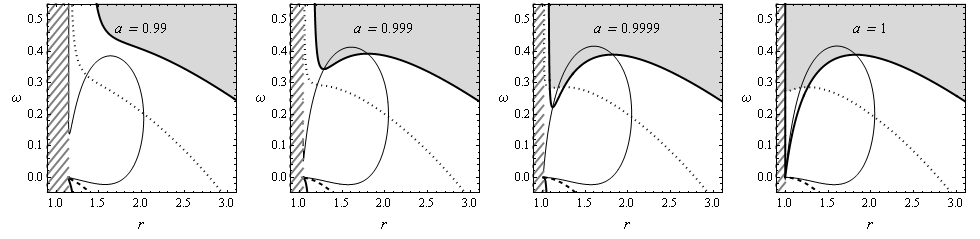}
\caption{\label{XXsitOM}
Properties of the fundamental angular frequency of the radial harmonic oscillations $\Omega_{\mir}(r;a,\omega)$ determined for near-extreme black holes with spin $a\in\{0.99,0.999,0.9999,1\}$ by the radial profiles of the characteristic functions $\omega(r;a)$. Regions where the string-loop motion is bounded and the expressions for the harmonic oscillations are relevant are greyed, regions below the  outer horizon are hatched. 
The dashed curves give the divergence function $\omega_{\rm d}(r;a)$, the dotted curves give the allowance function $\omega_{\rm a}(r;a)$, the thin curves give the extrema function $\omega_{\rm e(\mir)}(r;a)$. Thick black curves give the zero function $\omega_{\rm z(\mir)}(r;a)$. 
}
\end{figure*}
This is fulfilled when 
\beq
       \omega = \omega_{\rm z(\mir)}(r;a)
\eeq
where the function $\omega_{\rm z(\mir)}(r;a)$ is a proper solution of the equation 
\begin{widetext}
\bea
 \left(\omega ^2+1\right) \left[ G (r-1) \left(-3 a^2 r^2+6 a^2 r-6 a^2-5 r^4+12 r^3\right)- F G -2 F (a^2+3r^2) (r-1)\right] && \label{zIrI} \\
 +\frac{4 a \omega}{\sqrt{\Delta }} \left[G (a^2+3r^2) \left(\Delta -(r-1)^2\right)-6 \Delta G (r-1) r+2 \Delta  (a^2+3r^2)^2 (r-1)\right] &=& 0 .  \nn
\eea
\end{widetext}
The reality condition governing existence of the solution of (\ref{zIrI}) enables us to determine regions where the radial frequency of the string loop oscillations around the equatorial stable equilibrium positions is well defined. We have found two regions of stable equilibrium positions, and the related radial and vertical frequencies of the string loop oscillations; the additional inner region is located very close to the horizon in the field of near-extreme black holes with the spin $a>a_{\rm e(\mir)} \sim 0.9963$ that can occur for a limited range of the string loop parameter $\omega$. We demonstrate the special character of the behavior of the energy boundary function $E_{\rm b}(r;a,J,\omega)$ in the field of near-extreme Kerr black holes and related typical meridional sections of constant energy in Fig. \ref{fceEB}. 

\begin{figure}
\includegraphics[width=0.8\hsize]{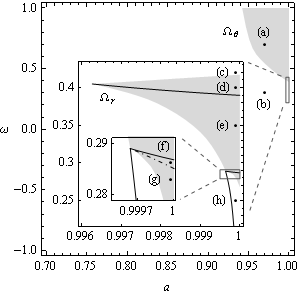}
\caption{\label{regionAW} 
Distribution of the various types of the behavior of the angular frequencies of radial and vertical string-loop harmonic oscillations in the parameter space $a$---$\omega$. The complex phenomena occur for positive values of the string-loop parameter $\omega$, and for the vertical frequency in spacetimes having spin $a>a_{\rm e(\mit)} \sim 0.942$, and for the radial frequency at near-extreme Kerr black-hole spacetimes with $a>a_{\rm e(\mir)} \sim 0.9963$. Representative points are depicted as (a)-(h) and the corresponding radial profiles of the radial and vertical angular frequencies are given in Figure \ref{exXX}.
}
\end{figure}

The local extrema of the radial profiles of the angular frequencies of the vertical oscillations are given by the condition 
\beq 
       \frac{\d \Omega_{\mit}(r;a,\omega)}{\d r} = 0 .
\eeq
This leads to vanishing of a very complex polynomial expression in all the variables $r,a,\omega$ and the solution in the form of the function $\omega_{\rm e(\mit)}(r;a)$ is illustrated for typical values of the black hole spin $a=0, 0.9, 1$ in Fig. \ref{TTsitOM} along with the functions $\omega_{\rm d}(r;a)$ and $\omega_{\rm a}(r;a)$. The local extrema are allowed for spin $a>0$ and for $\omega>0$. However, the physically relevant oscillations occur only for the string loop parameter $\omega$ allowing also the radial oscillations. Outside this region, the function $\Omega_{\mit}(r;a,\omega)$ governs oscillations of string loops falling into the black hole. The physically relevant extrema for string loops oscillating in vicinity of stable equilibrium positions can exist for the black hole spin $a > a_{\rm e(\mit)} \sim 0.942$. Note that relevance of the vertical (and radial) oscillations is restricted by the allowance function $\omega_{\rm a}(r;a)$. We can see in Figure \ref{TTsitOM} that the divergent and zero points of the angular frequency of vertical oscillations are forbidden by the allowance condition for all non-extreme Kerr black holes. In the case of extreme black holes ($a=1$), the zero point is located at $r=1$ where also the horizon is located, but this surface is degenerated, having infinite proper length, as well known \cite{Bar-Pre-Teu:1972:ApJ:}. 

\begin{figure*}
\includegraphics[width=\hsize]{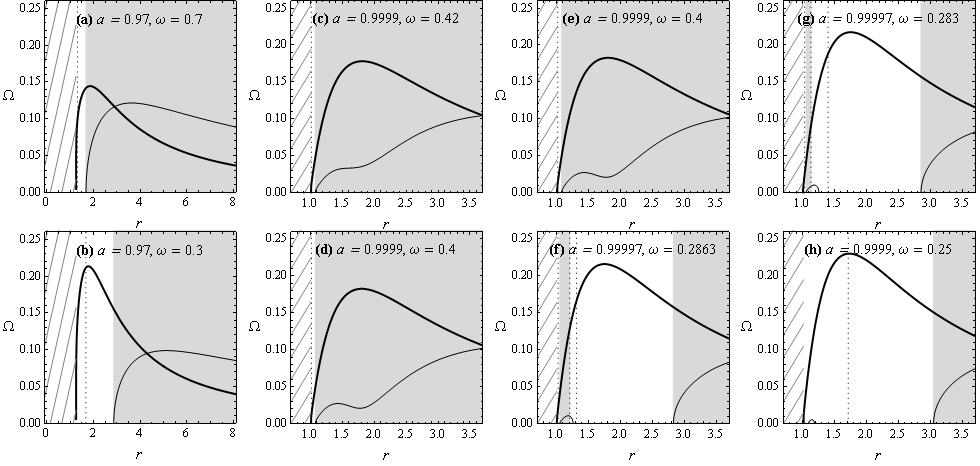}
\caption{\label{exXX}
The radial profiles of the angular frequencies of the string-loop harmonic oscillations $\Omega_{\mir}$ (thin) and $\Omega_{\mit}$ (thick) in the near-extreme Kerr black hole spacetimes. The profiles are given in the representative cases depicted in Figure \ref{regionAW} that cover all qualitatively differing cases; the spin and angular momentum parameters $a,\omega$ are given explicitly in the subfigures. The dotted lines give the allowance limit. Regions where the string-loop motion is bounded and the profiles are physically relevant are grayed, multiple extrema can exits for the radial angular frequency $\Omega_{\mir}$ being dependent on $\omega$ parameter.}
\end{figure*}

The local extrema of the radial profiles of the angular frequency of the radial oscillations are given by the condition  
\beq
    \frac{\d \Omega_{\mir}(r;a,\omega)}{\d r} = 0 .         
\eeq
From the asymptotic behavior of the radial profiles of the harmonic oscillations of the string loops and the vanishing of the $\Omega_{\mir}(r;a,\omega)$ at the innermost stable position, it is clear that at least one local maximum exists for the radial frequency profile. The implicit numerical solution of the extrema equation, given by the function $\omega_{\rm e(\mir)}(r;a)$, is illustrated for typical values of the black hole spin $a=0, 0.9, 1$ in Fig. \ref{RRsitOM}, along with the functions $\omega_{\rm z(\mir)}(r;a)$, $\omega_{\rm d}(r;a)$ and $\omega_{\rm a}(r;a)$.  The additional local extrema occur in the near-extreme black holes backgrounds with $a>a_{\rm e(\mir)} \sim 0.9963$. 

Using the characteristic functions of the parameter $\omega$ represented in figures \ref{TTsitOM} and \ref{RRsitOM}, we give important examples of the radial profiles of the string-loop fundamental frequencies $\Omega_{\mir}$ and $\Omega_{\mit}$ in figure \ref{exFF}. The regions where "lakes" of the energy boundary function $E_{\rm b} (r,\theta;a,J,\omega)$ can exist and the string motion is bounded (being of the type 3 on Fig. \ref{string_clasBH}) are explicitly marked. For the black hole spin, we chose typical values of $a=0, 0.9, 0.99, 1.0$. For the parameter $\omega$, we use three characteristic values, namely ($-1,0,1$), that properly represent the whole range of the parameter $\omega$. This covers all the space of parameters $a, \omega$, with exception of a small region discussed in detail later. 

While Fig. \ref{TTsitOM} fully reflects all qualitative aspects of the angular frequency of the vertical oscillations, Figure \ref{RRsitOM} representing the properties of the radial angular frequency $\Omega_{\mir}(r;a,\omega)$ is incomplete. The case of near-extreme black holes with $a \sim 1$ has to be treated separately.
The behavior of the characteristic functions of the parameter $\omega$, i.e., the functions $\omega_{\rm z(\mir)}(r;a)$, etc., represented in Fig. \ref{RRsitOM}, has to be completed for their detailed behavior in the case of near-extreme black holes with the spin $a$ taking values ($0.99,0.999,0.9999,1$) represented by Fig. \ref{XXsitOM}. Figs. \ref{RRsitOM} and \ref{XXsitOM} reflect all the qualitatively different possibilities of the behavior of the characteristic functions.
Notice that in the case of $a=0.9999$, and, of course, in the case $a=1$, the allowance condition $\omega > \omega_{\rm a}(r;a)$ directly restricts the region of the bounded oscillations. The allowance condition can be relevant in restricting the zero $\omega_{\rm z(\mir)}(r;a)$ and extrema $\omega_{\rm e(\mir)}(r;a)$ curves of the radial oscillatory frequency in the black hole spacetimes with $a>a_{\rm \mir(a)} \doteq 0.99964$. 

In the parameter space $a$---$\omega$, the distribution of qualitatively different types of the behavior of the radial profiles of the angular frequency $\Omega_{\mit}$ ($\Omega_{\mir}$) having local extrema (additional local extrema or additional zero points) is presented in Figure \ref{regionAW}. For near-extreme black holes, the typical cases of the $\Omega_{\mit}$ and $\Omega_{\mir}$ radial profiles are represented in Figure \ref{exXX} for the parameters ($a,\omega$) corresponding to the points (a)-(h) depicted in Figure \ref{regionAW}. According to the value of the string loop parameter $\omega$, we can distinguish the situations where there are three local extrema, two maxima and one minimum, or two maxima and two additional zero points, or the restricted situations where the internal region of the oscillatory motion allows only segments of the region between two internal zero points of $\Omega_{r}$ -- see Figure \ref{exXX}.

\begin{figure*}
\includegraphics[width=\hsize]{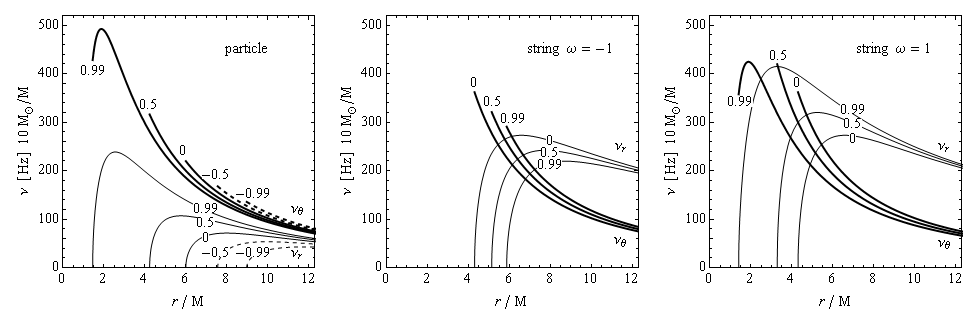}
\caption{\label{psQPO}
Test-particle epicyclic frequencies (left), and string-loop oscillatory frequencies $\nu_\mir, \nu_\mit$ are given for the minimal string-loop angular momentum parameter $\omega = -1$ (middle) and the maximal angular momentum parameter $\omega = 1$ (right) for increasing spin $a$ in the field of Kerr black holes with $M = 10 M_{\odot}$. The radial profiles of the vertical (radial) frequencies are given by thick (thin) lines. The vertical frequency curves are restricted to the region of existence (zero point) of the corresponding radial frequency curves.
Note that there is $(a\omega)\leftrightarrow-(a\omega)$ symmetry for string-loop oscillatory motion, while it is not present for the test-particle epicyclic motion.)
}
\end{figure*}

The coincidence condition for the radial and vertical frequencies of the string loop oscillations reads 
\beq
       \Omega_\mir(r;a,\omega) = \Omega_{\mit}(r;a,\omega) .       
\eeq
We have found that there is only one coincidence radius for all values of the spin $a$ and the rotational string loop parameter $\omega$ in the Kerr black hole spacetimes. Notice that the situation can be more complex in the Kerr naked singularity spacetimes, similarly to the case of the test particle epicyclic motion \cite{Stu-Sche:2012:CLAQG:}. 

\subsection{Classification of the string loop oscillations in the black hole spacetimes}

We give classification of the radial profiles of the string oscillatory motion relative to the spin parameter $a$ of the black hole and the angular momentum parameter $\omega$ of the string loop, focusing attention only on the properties of the oscillatory motion in the regions when the motion is bounded and both the angular frequencies are well defined simultaneously. There are four types of the behavior when at least one of the angular frequencies demonstrates different properties. We give properties of the radial profiles in all the four types.

{\it Type I}

The radial angular frequency profile has one zero point and one local maximum. The vertical angular frequency is monotonously decreasing function of radius.

{\it Type II}

The radial angular frequency profile has one zero point and one local maximum. The vertical angular frequency profile has one local maximum and the zero point at the horizon; the maximum is located in the physically relevant region of simultaneous relevance of the radial angular frequency, the zero point is located in the "unphysical" region. 

{\it Type III}

The radial angular frequency profile has two segments. The outer one has a local maximum and a zero point, the inner one has a local maximum in between two zero points; for near-extreme Kerr black holes with $a>a_{\rm \mir(a)}$ only part of the inner segment is relevant. The extension of the region of relevance depends on the parameter $\omega$ as demonstrated in Figs. \ref{exXX}f-h.
The vertical frequency profile has a zero point at the horizon (physically irrelevant) and a local maximum, but only two segments of the profile are relevant, corresponding to the regions where also the radial angular frequency is well defined; the inner segment is increasing, while the outer segment is decreasing with radius increasing. 

{\it Type IV}

The radial angular frequency profile has one zero point, two local maxima and one local minimum in between them. The vertical angular frequency profile has a zero point at the horizon and a local maximum. The physically relevant parts correspond to the regions where both the frequencies are defined. 
 
In the space of parameters $a-\omega$, we can define classes of the Kerr black hole spacetimes according to the types of the radial profiles of the radial and vertical angular frequencies of the string oscillations allowed in them. The classification is governed by Figure \ref{regionAW}. 

\begin{figure*}
\includegraphics[width=\hsize]{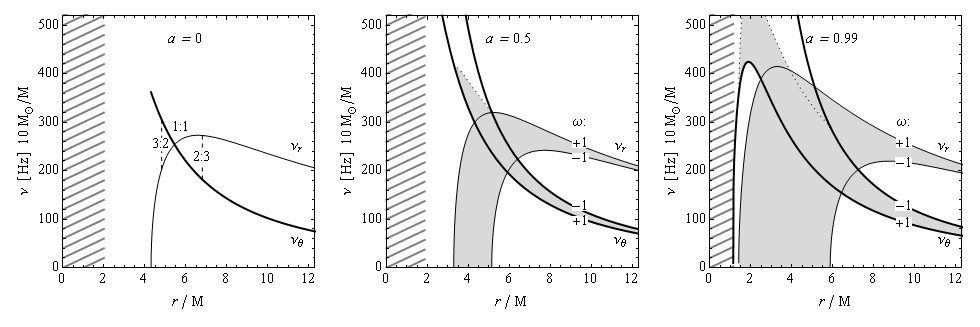}
\caption{\label{exQPO}
Extension of the radial profiles of the angular frequency of the radial and vertical harmonic oscillation of string loops given for the complete range of the string loop parameter $\omega$. The extension is illustrated for typical values of the black holes spin $a=0,0.5,0.99$. The relevant region where both vertical and radial frequencies with the same parameters are defined simultaneously is greyed.
}
\end{figure*}

We have to stress that for the near-extreme black holes with spin $a>a_{\rm e(\mir)} \sim 0.9963$, we have found a special "strange" behavior of the radial oscillatory frequency with a small hump in the radial profile that can be transformed to the case when a separate additional "hill" occurs in the radial profile. It is of some interest that a similar strange 'humpy' behavior occurs in the structure of the LNRF velocity profiles of the toroidal configurations of perfect fluid in the field of near-extreme black holes with spin $a > 0.9999$ \cite{Stu-Sla-Tor-Abr:2005:PHYSR4:}. This effect is connected to the so called Aschenbach effect demonstrating a humpy structure of the LNRF velocity profiles of Keplerian discs and occurring in the field of black holes with spin $a>a_{\rm A}=0.983$ \cite{Asch:2004:ASTRA:}. It could be interesting to look for a possible relation of the special 'humpy' effect of the string-loop oscillations and the Aschenbach effect of the geodetical motion related to properties of the perfect fluid toroidal configurations.

\section{String-loop oscillations at resonance radii as a model of twin HF QPOs in microquasars \label{observations}}

The quasi-periodic character of the motion of string loops trapped in a toroidal space around the equatorial plane of a Kerr black hole (naked singularity) suggests interesting astrophysical application related to the HF QPOs observed in binary systems containing a black hole or a neutron star, or in active Galactic nuclei. Some of the~HF~QPOs come in pairs of the~upper and lower frequencies ($\nu_{\mathrm{U}}$, $\nu_{\mathrm{L}}$) of {\it twin peaks} in the~Fourier power spectra. Since the~peaks of high frequencies are close to the~orbital frequency of the~marginally stable circular orbit representing the~inner edge of Keplerian discs orbiting black holes (or neutron~stars), the~strong gravity effects must be relevant in explaining HF~QPOs \citep{Tor-etal:2005:ASTRA:}. Usually, the Keplerian orbital and epicyclic (radial and latitudinal) frequencies of geodetical circular motion \cite{Tor-Stu:2005:ASTRA:,Kot-Stu-Tor:2008:CLAQG:,Stu-Kot:2009:GenRelGrav:} are assumed in models explaining the HF QPOs in both black hole and neutron star systems. 

Before the~twin peak HF~QPOs have been discovered in microquasars \citep[first by][]{Str:2001:ASTRJ2L:}, and the~3\,:\,2 ratio pointed out, \citep{Klu-Abr:2000:ASTROPH:} suggested on theoretical grounds that these QPOs should have rational ratios, because of the~resonances in oscillations of nearly Keplerian accretion disks; see also \citep{Ali-Gal:1981:GENRG2:}. It seems that the~resonance hypothesis is now well supported by observations, and that the~3\,:\,2 ratio ($2\nu_{\mathrm{U}} = 3\nu_{\mathrm{L}}$) is seen most often in twin peak QPOs in the~LMXB containing black holes (microquasars). In addition, there is even some evidence of the~same 3\,:\,2 ratio in the~X-ray spectra of the~Galaxy centre black hole in Sgr\,A$^*$ \citep{Abr-etal:2004:ASTRJ2L:,Asch:2004:ASTRA:,Ter:2005:ASTRA:}, the~Galactic nuclei MCG-6-30-15 and NGC 4051 \citep[]{Lac-Cze-Abr:2006:astro-ph0607594:}, and other active Galactic nuclei \citep{Mid-Utt-Don:2011:,Mid-Don-etal:2009:}. Here we concentrate on the case of three microquasars, GRO 1655-40, XTE 1550-564 and GRS 1915+105, that are widely discussed in recent literature \cite{Tor-etal:2011:ASTRA:}.  

Unfortunately, neither of the recently discussed models is able to explain the HF QPOs in all the microquasars \cite{Tor-etal:2011:ASTRA:}. Therefore, it is of some relevance to let the string loop oscillations, characterized by their radial and vertical (latitudinal) frequencies, to enter the play, as these frequencies are comparable to the epicyclic geodetical frequencies, but slightly different, enabling thus some relevant corrections to the predictions of the models based on the geodetical epicyclic frequencies $\nu_\mit, \nu_\mir$. We again keep the assumption of the resonance phenomena occurring in the oscillatory motion. The resonant phenomena (parametric or forced) are discussed in standard textbooks \cite{Lan-Lif:1976:Mech:,Nay-Moo:1979:NonOscilations:}, discussion of their relevance to the accretion phenomena can be found, e.g., in \cite{Stu-Kot-Tor:2013:ASTRA:}. For string-loop small oscillations, the resonance phenomena are governed by the KAM theory \cite{Arnold:1978:book:}. Here we assume their relevance, postponing a detailed discussion to a future work. 

\begin{figure*}
\includegraphics[width=\hsize]{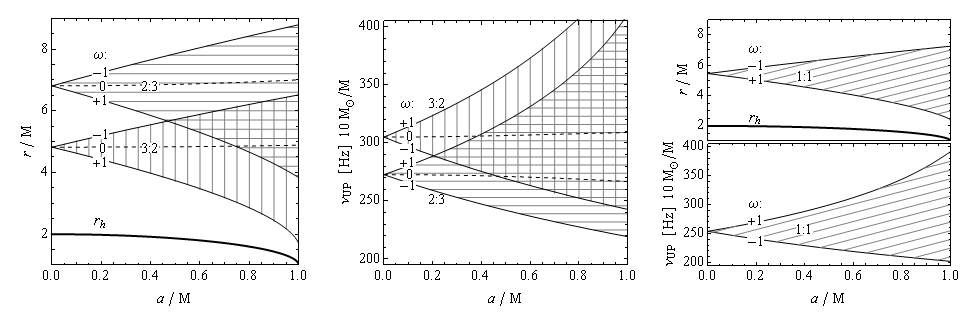}
\caption{\label{rezRR}
(Left) The inner and outer resonant radii where the radial and vertical frequencies of the string-loop oscillations have rational ratio $3:2$ or $2:3$, being dependent on the Kerr spin parameter $a$ and the complete range of the string-loop parameter $\omega$; the case of $\omega=0$ is given by dotted lines. The loci of the outer horizon are also presented. (Middle) Extension of the related upper resonant frequency occurring at the inner and outer resonant radii are given as a function of the black-hole spin $a$, for the whole range of the string-loop parameter $\omega$; the case of $\omega=0$ corresponds to dotted lines. For the $3:2$ ($2:3$) ratio the upper frequency is the one corresponding to the vertical (radial) oscillations. (Right) The resonant "coincidence" radii with the frequency ratio $1:1$ and related values of the common frequency of the radial and vertical oscillation are illustrated for comparison.}
\end{figure*}

We can assume applicability of the parametric resonance, discussed in \cite{Lan-Lif:1976:Mech:}, focusing attention to the case of the frequency ratios $\nu_{\mit}:\nu_{\mir} = 3:2$ or $\nu_{\mit}:\nu_{\mir} = 2:3$, as the observed values of the twin HF QPO frequencies for GRO 1655-40, XTE 1550-564 and GRS 1915+105 sources show clear ratio 
\beq
 \nu_{\rm U} : \nu_{\rm L} = 3 : 2
\eeq
for the upper $\nu_{\rm U}$ and lower $\nu_{\rm L}$ frequencies - see Tab.\ref{tab1}. We identify directly the frequencies $\nu_{\rm U}, \nu_{\rm L}$ with $\nu_{\mit}, \nu_{\mir}$ or $\nu_{\mir}, \nu_{\mit}$  frequencies. In contrast to the resonance epicyclic model, the string loop oscillation model allows both frequency ratios
\beq
  \nu_{\mit} : \nu_{\mir} = 3 : 2, \quad \nu_{\mit} : \nu_{\mir} = 2 : 3.
\eeq
Since $ r_{3:2} < r_{2:3} $, we call the first resonance radius, where $\nu_\mit : \nu_{\mir} = 3 : 2$, the inner one, and the second resonance radius, where $\nu_\mit : \nu_{\mir} = 3 : 2$, the outer one. Note that for the oscillating string loops also the $1:1, 1:2, 2:1$, or other, resonant frequency ratios can be relevant - these can be relevant in other sources where such frequency ratios are observed, see, e.g., \cite{Lac-Cze-Abr:2006:astro-ph0607594:}, and will be discussed in future works. 

We have to stress that the region of the innermost oscillations occurring in close vicinity of the horizon in the field of near-extreme black holes with $a>a_{\rm e(\mir)}=0.9963$ can be excluded from our discussion of the oscillations with the frequency ratio $3:2$ (or $2:3$), since we can demonstrate that the frequency ratios occurring in those regions are substantially higher (lower)- see Fig. \ref{exXX}. 

The frequency of the radial oscillations in the innermost regions of the near-extreme Kerr black holes is substantially lower in comparison with the frequency of the vertical oscillations, opening up  potentially the possibility to relate them to the low-frequency QPOs observed sometimes in the microquasars along with the HF QPOs. However, such a possibility has to be studied very carefully, because the effective potential of such oscillatory motion is very shallow and strongly restricted to close vicinity of the innermost stable equilibrium points, as demonstrated in Figure \ref{fceEB}; for a similar situation related to the epicyclic test-particle motion see \cite{Stu-Kot-Tor:2011:ASTRA:}. Here, we focus our attention only to the oscillations around the outer stable equilibrium points where the frequency ratios $3:2$ can occur.

For characteristic values of the black hole spin $a=0,0.5,0.99$, we demonstrate behavior of the radial profiles of the radial and vertical frequency of the string-loop oscillations for the limiting values of the string-loop parameter $\omega = -1, 0, 1$ in Figure \ref{psQPO}, where comparison to the  frequencies of the epicyclic test-particle motion is also given. The qualitative differences are clearly seen, the crucial one being given by the fact that for the epicyclic test-particle motion there is always $\nu_{\mit}>\nu_{\mir}$, while for the string-loop oscillation frequencies there is $\nu_{\mit}<\nu_{\mir}$ at large radii, while $\nu_{\mit}>\nu_{\mir}$ at small radii. 

\begin{figure*}
\includegraphics[width=\hsize]{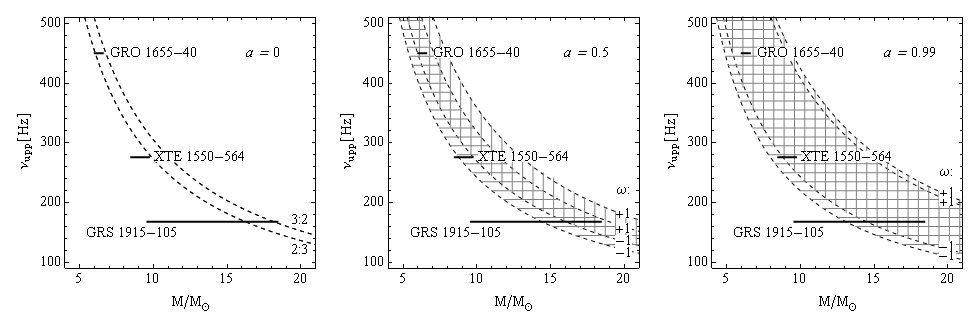}
\caption{\label{QPOfit}
The upper string-loop oscillation frequency $\nu_{\rm U}$ at the 3:2 or 2:3 resonance radii, calculated in the framework of the string-loop model with maximal range of the string-loop parameter $\omega$ as a function of the black hole mass for typical values of the black hole spin $a=0,0.5,0.99$, and compared to the mass-limits obtained from observations of the three microquasars GRO 1655-40, XTE 1550-564, GRG 1915-105. Hatched areas cover the whole interval of $\omega\in\langle-1,1\rangle$. The vertical hatch corresponds to the $3:2$ frequency ratio $\nu_{\mit}/\nu_{\mir}$, while the horizontal hatch corresponds to the $2:3$ frequency ratio. The frequency $\nu_{\rm U}$ can appear in both 3:2 or 2:3 resonance radii. In the common areas both 3:2 or 2:3 frequency ratios are possible, but for different $\omega$ parameters.
}
\end{figure*}

\begin{table}[!h]
\begin{center}
\begin{tabular}{c l l l}
\hline
Source & GRO 1655-40 & XTE 1550-564 & GRS 1915+105 \\
\hline \hline
$ \nu_{\rm U}$ [Hz] & $447${\linka}$453$ & $273${\linka}$279$ & $165${\linka}$171$ \\
$ \nu_{\rm L}$ [Hz] & $295${\linka}$305$ & $179${\linka}$189$ & $108${\linka}$118$ \\
$ M/M_\odot $ & $6.03${\linka}$6.57$ & $8.5${\linka}$9.7$ & $9.6${\linka}$18.4$ \\
$ a $ & $0.65${\linka}$0.75$ & $0.29${\linka}$0.52$ & $0.98${\linka}$1$ \\
\hline
\end{tabular}
\caption{Observed twin HF QPO data for the three microquasars, and the restrictions on mass and spin of the black holes located in them, based on measurements independent on the HF QPO measurements. The restrictions on the mass and spin are reflected by the gray rectangulars.} \label{tab1}
\end{center}
\end{table}
\begin{table}[!h]
\begin{center}
\begin{tabular}{c c c c}
\hline
Source & GRO 1655-40 & XTE 1550-564 & GRS 1915+105 \\
\hline \hline
$ \omega_{3:2} $ & $-0.42${\linka}$-0.09$ & $-1$ & $-1${\linka}$0.01$ \\
$ \omega_{2:3} $ & $0.01${\linka}$0.25$ & $-1${\linka}$-0.05$ & $-1${\linka}$0.26$ \\
$ M/M_\odot $ & $6.03${\linka}$6.57$ & $8.7${\linka}$9.7$ & $13.1${\linka}$18.4$ \\
\hline
\end{tabular}
\caption{Restrictions on the $\omega$ parameter of the string-loop resonance oscillation model of twin HF OPOs obtained for the three microquasars. The limits on the black hole mass implied by the string-loop model are presented. No additional limits on the spin parameter $a$ have been obtained.} \label{tab2}
\end{center}
\end{table}

Further we demonstrate extension of the radial profiles of the frequencies of the string loop oscillations for the typical values of the black holes spin $a=0,0.5,0.99$ in Figure \ref{exQPO}. The extension of the radial profiles covering the whole range of $-1 < \omega < 1$ increases strongly with the spin increasing. The extension is vanishing in the degenerate case of the Schwarzschild spacetimes with $a=0$ and grows more strongly for positive values of the string-loop angular momentum parameter $\omega$. Of course, this effect extends the possibility to meet the observationally given frequencies in the microquasars by the string-loop oscillation frequencies with increasing spin of the black hole background. 

The observed twin HF QPOs are determined by the frequency ratio and one of the observed frequencies -- we use the upper of the observed frequencies (here, we do not consider the other measurable details of the observed oscillations, focusing attention on their frequencies only). The frequency ratio determines the resonant radius and the upper frequency, if the black hole spin $a$ and the string loop parameter $\omega$ are given. We illustrate the dependence of the resonant radii and the related upper frequency on the parameters $a$ and $\omega$ for both $3:2$ and $2:3$ ratios of $\nu_{\mit}:\nu_{\mir}$ in Figure \ref{rezRR}. We can see that the resonant radii $r_{3:2}$ and $r_{2:3}$ are separated for spin $a<0.5$, while the upper frequencies $\nu_{\rm U(3:2)}=\nu_{\mit}$ and $\nu_{\rm U(2:3)}=\nu_{\mir}$ are separated for $a<0.2$. Notice that the resonant radii $r_{3:2}(\omega=0)$ and $r_{2:3}(\omega=0)$, and the resonant upper frequencies $\nu_{\rm U(3:2)}(\omega=0)$ and $\nu_{\rm U(2:3)}(\omega=0)$, depend only very weakly on the spin parameter $a$. 

\begin{figure*}
\includegraphics[width=\hsize]{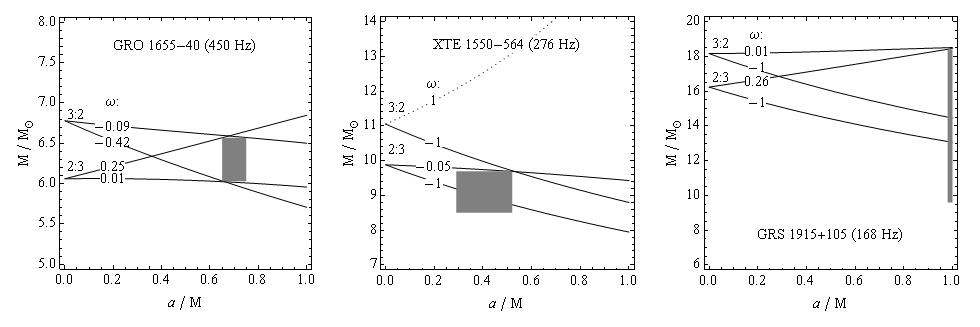}
\caption{Restrictions on the parameter $\omega$ of the string-loop oscillatory model of HF QPOs,  obtained for the three microquasars GRO 1655-40, XTE 1550-564, GRG 1915-105 due to the limits on their mass and spin obtained by observations independent on the HF QPO observations. The limits are given by the shaded rectangulars. \label{aMctvereckyBH}}
\end{figure*}

Now we can turn to the possibility to fit the observed data in the three microquasars, GRO 1655-40, XTE 1550-564 and GRS 1915+105, to the string-loop resonance oscillation model.  

For fixed black hole spin $a$ and fixed string loop parameter $\omega$, the upper frequency of the twin HF QPOs can be given as a function of the black hole mass $M$. If the black hole mass is restricted by independent observations, as is usually the case, we can obtain some restrictions on the string-loop resonant oscillations model, as illustrated in Figure \ref{QPOfit}., where the situation is demonstrated for typical values of the spin $a = 0, 0.5, 0.99$, and the limits on the black hole mass, given in Tab. \ref{tab1}, are applied. We can see immediately that for the Schwarzschild black hole ($a=0$), the string loop model can explain only the HF QPOs in the source GRS 1915+105, while introducing the spin and parameter $\omega$ improves the applicability of the string-loop resonant oscillations model substantially and enables explanation of the twin HF QPOs in all three microquasars. moreover, the string-loop resonant oscillations model predicts HF QPO frequencies that put generally additional restrictions on the black hole parameters, as illustrated in Figure \ref{QPOfit}. 

The strongest predictions are possible, if we use also the independent restrictions on the dimensionless spin $a$ of the black holes that were obtained in the case of the three microquasars by the spectral fitting method (see \cite{McCli-Rem:2011:CLAQG:}), and are presented in Table \ref{tab2}. Using the upper frequencies at the resonant radii in the framework of the string-loop resonant oscillations model, we obtain the relations $M_{3:2}=M_{3:2}(a;\omega)$ and $M_{2:3}=M_{2:3}(a;\omega)$ that can effectively restrict the range of mass parameter $M$ for a fixed spin $a$ and the whole range of the string parameter $\omega$. The cases $\omega= +1, -1$ determine the limiting curves in the $M-a$ diagram as demonstrated in Fig. \ref{aMctvereckyBH}. They can be confronted with the observational independent limits on the black hole mass and dimensionless spin, given by the shaded rectangulars in figure \ref{aMctvereckyBH}. In such a way, we clearly demonstrate that the string-loop resonant oscillations model can successfully explain the frequencies of the twin HF QPOs observed in the three considered microquasars. 

The observational results obtained for the three microquasars, giving restrictions on the mass and spin of their central black holes, constrain substantially the string-loop resonant oscillations model as demonstrated in Table \ref{tab2}, putting relevant limits on the angular momentum parameter $\omega$ of the string loop. In two of the sources, GRO 1655-40, and GRS 1915+105, both the ratios $\nu_{\mit} : \nu_{\mir} = 3:2, 2:3$ are allowed, while for the source XTE 1550-564, only the ratio $\nu_{\mit} : \nu_{\mir} = 2:3$ is allowed. The string-loop resonant oscillations model puts also strong limits on the values of the string-loop parameter $\omega$ for all three microquasars. The data obtained for the three microquasars and the limits on their mass and spin determined by observations independent on the HF QPO measurements (e.g., the spectral fitting restricting the spin) imply that the string-loop resonance oscillations model with negative values of the parameter $\omega$ is preferred, with one exception of the oscillations at the resonant point $r_{2:3}$ of the microquasar GRO 1655-40 where only positive values of $\omega$ are allowed -- see Table \ref{tab2}. A detailed information is contained in figure \ref{aMctvereckyBH}. We can generally say that for a given box of values of mass and spin of a black hole, the resonance point at $r_{3:2}$ requires values of the string loop parameter $\omega$ shifted to lower values than the resonance point at $r_{2:3}$.

The string-loop resonant oscillations model of the HF QPOs implies also additional restrictions on the mass and dimensionless spin of the central black hole in the microquasars. In the case of the three microquasars GRO 1655-40, XTE 1550-564, GRG 1915-105, we have found no additional restriction on the range of the mass parameter in the microquasar GRO 1655-40, a small restriction of the mass parameter of the microquasar XTE 1550-564, while in the case of the microquasar GRS 1915+105, the model admits only mass of the black hole $M > 13 M_{\odot}$ that represents a large reduction of the original restriction. No additional restriction on the black hole spin parameter is obtained in the case of all three microquasars. 

\section{Conclusions}

We have studied frequencies of harmonic radial and vertical oscillations of relativistic current-carrying string loops in the equatorial plane of Kerr black holes, comparing them to the analogous frequencies of the radial and vertical geodetical epicyclic oscillatory motion of test particles. Magnitudes of both the string loop and test particle frequencies are comparable, having the same mass scaling, however, their radial profiles differ significantly. While the radial epicyclic frequency is always smaller than the vertical epicyclic frequency, the frequencies of the radial and vertical oscillations of the string loops coincide at a "coincidence" radius; the radial frequency overcomes the vertical frequency at radii exceeding the coincidence radius and is smaller under the coincidence radius. 

The string loop oscillations could be related to the twin HF QPOs observed in microquasars, if we assume their occurrence at resonant radii, where the ratio of the frequencies becomes rational, and a parametric resonance can be relevant. In such a case, we have to consider a dichotomy introduced by the behaviour of the radial profiles of the radial and vertical string loop oscillations, since two resonant points with the same frequency ratio arise for string loop oscillations. In the field of Kerr black holes, the frequencies of the harmonic oscillations of string loops depend on the parameter $\omega$ belonging to the interval $\langle-1,1\rangle$. that reflects the axial angular momentum parameter combining effects of both the angular momentum and tension related to the string, and representing a constant of the string-loop motion. The role of the $\omega$ parameter increases with increasing spin of the black hole, however, for the Schwarzschild spacetime ($a=0$) degenerate situation arises, as the string loop motion is independent of $\omega$. Note that the interplay of the black hole spin $a$ and the string loop parameter $\omega$ has a crucial role in explaining the HF QPOs as the range of allowed frequencies governed by the whole interval of allowed values of $\omega$ increases strongly with increasing spin (see Figs. \ref{psQPO}. and \ref{QPOfit}.) extending thus the ability to cover the observed HF QPO phenomena. 

In the quasi-periodic regime the frequencies of the string-loop harmonic oscillations are relevant \cite{Stu-Kol:2012:JCAP:,Kol-Stu:2013:PHYSR4:}, therefore, we applied the model of the string-loop oscillations resonant at radii corresponding to the frequency ratio $\nu_{\mit} : \nu_{\mir} = 3:2, 2:3$ to the twin HF QPOs observed with frequency ratio $\nu_{\rm U} : \nu_{\rm L} = 3:2$ in three microquasars GRO 1655-40, XTE 1550-564 and GRS 1915+105. We have demonstrated that the model of the string-loop resonant oscillations can well explain the observed twin HF QPOs for whole the range of the spin parameters expected in the three microquasars and implies a restriction of the range of the mass parameter in microquasars XTE 1550-564 and GRS 1915+105. The observational data put significant restrictions on the string-loop parameter $\omega$ in the case of all three microquasars. The string-loop resonant oscillations model can thus suffice a satisfactory explanation of the twin HF QPOs observed in microquasars, and can give an additional restriction on the black hole parameters. 

The radial profile of the radial and vertical string-loop oscillations, demonstrating also the case of a unique observed frequency occurring at the coincidence radius, enables potentially explanation of more complex frequency patterns observed in some microquasars, or even in some binary systems containing a neutron star. The test of applicability of the string-loop resonant oscillations model in the cases when more than two HF QPOs are observed, as is the case of the GRS 1915+105 microquasar, or of some other sources, is of high relevance. Such situations will be studied in future works. 

We conclude that the model of string-loop resonant oscillations can be relevant for the explanation of the HF QPOs observed in microquasars and binary systems containing a neutron (quark) star, and it deserves further investigation related to a more detailed physical description of the properties of the string loops and their oscillations, and models of optical phenomena related to radiating string
loops in oscillatory motion.

\section*{Acknowledgments}

The authors would like to acknowledge the internal student grant of the Silesian University SGS/23/2013, the EU Synergy Grant No. CZ.1.07/2.3.00/20.0071, and the GA\v{C}R excellence grant 14-37086G for support.





\begin{thebibliography}{99}

\bibitem{Jac-Sot:2009:PHYSR4:}
   T.~Jacobson and T.~P.~Sotiriou, 
   ``String dynamics and ejection along the axis of a spinning black hole,``
   Phys.\ Rev.\ D {\bf 79}, 065029, (2009)
   [arXiv:gr-gc/0812.3996].
   
\bibitem{Kol-Stu:2013:PHYSR4:}
 	M.~Kolo{\v s} and Z.~Stuchl{\'{\i}}k,
  ``Dynamics of current-carrying string loops in the Kerr naked-singularity and black-hole spacetimes,''
  submitted to Phys.\ Rev.\ D {\bf 88}, 065004 (2013).
	
\bibitem{Kol-Stu:2010:PHYSR4:}
 	M.~Kolo{\v s} and Z.~Stuchl{\'{\i}}k,
  ``Current-carrying string loops in black-hole spacetimes with a repulsive cosmological constant,''
  Phys.\ Rev.\ D {\bf 82}, 125012 (2010)
	[arXiv:gr-qc/1103.40056].

\bibitem{Stu-Kol:2012:PHYSR4:}
 	Z.~Stuchl{\'{\i}}k and M.~Kolo{\v s},
  ``Acceleration of string loops in the Schwarzschild-de~Sitter geometry,''
  Phys.\ Rev.\ D {\bf 85}, 065022 (2012)
	[arXiv:gr-qc].

\bibitem{Lar:1994:CLAQG:}
  A.~L.~Larsen,
  ``Chaotic string capture by black hole,''
  Class.\ Quant.\ Grav.\  {\bf 11}, 1201 (1994)
  [arXiv:hep-th/9309086].
  
\bibitem{Fro-Lar:1999:CLAQG:}
  A.~V.~Frolov and A.~L.~Larsen,
  ``Chaotic scattering and capture of strings by a black hole,''
  Class.\ Quant.\ Grav.\  {\bf 16}, 3717 (1999)
  [arXiv:gr-qc/9908039].
	 
\bibitem{Sem-Dya-Pun:2004:Sci:}
  V.~Semenov, S.~Dyadechkin and B.~Punsly,
  ``Simulations of jets driven by black hole rotation,''
  Science {\bf 305}, 978 (2004)
  [arXiv:astro-ph/0408371].

\bibitem{Chri-Hin:1999:PhRvD:}
  M.~Christensson and M.~Hindmarsh,
  ``Magnetic fields in the early universe in the string approach to MHD,''
  Phys.\ Rev.\  D {\bf 60}, 063001 (1999)
  [arXiv:astro-ph/9904358].
  
\bibitem{Sem-Ber:1990:ASS:}
	V.~S.~Semenov and L.~V.~Bernikov, 
  ``Magnetic flux tubes - nonlinear strings in relativistic magnetohydrodynamics,''
  Astro.\ and Space Sci.\ {\bf 184}, 157-166 (1991).

\bibitem{Spr:1981:AA:}
	H.~C.~Spruit, 
	``Equations for thin flux tubes in ideal MHD,''
	Astron.\ Astrophys.\ {\bf 102}, 129 (1981).


\bibitem{Cre-Stu:2013:PhRvE:}
  C.~Cremaschini and Z.~Stuchl{\'{\i}}k,
  ``Magnetic loops generation by collisionless gravitationally-bound plasmas in axisymmetric tori''
  Phys.\ Rev.\ E {\bf 87}, 043113 (2013).

\bibitem{Arm-etal:2013:PRD:}
  A.~Tursunov, M.~Kolo{\v s}, B.~Ahmedov and Z.~Stuchl{\'{\i}}k,
  ``Dynamics of an electric current carrying string loop near a Schwarzschild black hole embedded in an external magnetic field,''  
  Phys.\ Rev.\ D {\bf 87}, 125003 (2013).

\bibitem{Stu-Kol:2012:JCAP:}
 	Z.~Stuchl{\'{\i}}k and M.~Kolo{\v s},
  ``String loops in the field of braneworld spherically symmetric black holes and naked singularities''
  JCAP, {\bf 10}, 008 (2012)

\bibitem{Bla-Zna:1977:MNRAS:}
  R.~D.~Blandford and R.~L.~Znajek,
  ``Electromagnetic extraction of energy from Kerr black holes,''
  Mon.\ Noti.\ of the Royal Astro.\ Society {\bf 179}, 433-456 (1977).

\bibitem{Gar-etal:2010:ASTRA:}
	J.~Gariel, M.~A.~H.~MacCallum, G.~Marcilhacy and N.~O.~Santos,
  ``Kerr geodesics, the Penrose process and jet collimation by a black hole,''
	Astronom.\ Astrophys.\ {\bf 515}, A15 (2010).

\bibitem{Pach-etal:2012:ApJ:}
 L.~A.~Pach{\'o}n, J.~A.~Rueda and C.~A.~Valenzuela-Toledo,
``On the Relativistic Precession and Oscillation Frequencies of Test Particles around Rapidly Rotating Compact Stars,''
Astro.\ Journal, {\bf 756}, 82 (2012),
[astro-ph.SR/1112.1712]. 

\bibitem{Gar-etal:2013:ApJ:}
	J.~Gariel, G.~Marcilhacy and N.~O.~Santos,
  ``Unbound Geodesics from the Ergosphere and the Messier 87 Jet Profile,''
	The Astrophysical Journal {\bf 774}, 109 (2013),
	[astro-ph.HE/1303.6474].

\bibitem{Stu:1983:BAC:}
  Z.~Stuchl{\'i}k,
  ``The motion of test particles in black-hole backgrounds with non-zero cosmological constant,''
  Bull.\ Astron.\ Inst.\ Czechosl.\ {\bf 34}, 129 (1983).
  
\bibitem{Stu-Hle:1999:PHYSR4:}
  Z.~Stuchl{\'i}k and S.~Hled{\'i}k,
  ``Some properties of the Schwarzschild--de~Sitter and Schwarzschild--anti-de~Sitter spacetimes,''
  Phys.\ Rev.\ D {\bf 60}, 044006 (1999).

\bibitem{Fer-etal:2012:ExpAstr:}
  M.~Feroci et~al, ``The Large Observatory for X-ray Timing (LOFT),''
	Experimental Astronomy {\bf 34}, 415-444 (2012),
	[astro-ph.IM/1107.0436].


\bibitem{Bar-Oli-Mil:2005:MONNR:}
	D.~Barret, J.~F.~Olive and  M.~C.~Miller,
  ``An abrupt drop in the coherence of the lower kHz quasi-periodic oscillations in 4U 1636-536,''
  Monthly Notices of the Royal Astronomical Society {\bf 361}, 855-860 (2005).

\bibitem{Bel-Men-Hom:2007:MONNR:BriNSQPOCor}
	T.~Belloni, M.~M{\'e}ndez and J.~Homan,
  ``On the kHz QPO frequency correlations in bright neutron star X-ray binaries,''
  Monthly Notices of the Royal Astronomical Society {\bf 376}, 1133-1138 (2007),

\bibitem{Kli:2000:ARASTRA:}  
  M.~van der Klis, ``Millisecond Oscillations in X-ray Binaries,''
  Annual Review of Astronomy and Astrophysics {\bf 38}, 717-760 (2000).

\bibitem{Rem:2005:ASTRN:}
	R.~A.~Remillard,
  ``X-ray spectral states and high-frequency QPOs in black hole binaries,''
  Monthly Notices of the Royal Astronomical Society {\bf 326}, 804-807 (2005),
  [astro-ph/0510699].
 
\bibitem{Rem-McCli:2006:ARASTRA:}
	 R.~A.~Remillard and J.~E.~McClintock,
	``Active X-ray States of Black Hole Binaries: Current Overview,''
  Bulletin of the American Astronomical Society {\bf 38}, 903 (2006),
  [astro-ph/0510699]. 
 
\bibitem{McCli-Rem:2004:CompactX-Sources:}
  J.~E.~McClintock and R.~A.~Remillard, ``Black Hole Binaries,''
  Compact Stellar X-Ray Sources, Cambridge University Press, Cambridge (2004),
  [astro-ph/0306213].
  
\bibitem{Tor-etal:2005:ASTRA:}
	G.~T{\"o}r{\"o}k, M.~A.~Abramowicz, W.~Klu{\'z}niak and Z.~Stuchl{\'i}k,
  ``The orbital resonance model for twin peak kHz quasi periodic oscillations in microquasars,''
	Astronom.\ Astrophys.\ {\bf 436}, 1-8 (2005).
	  
\bibitem{McCli-etal:2006:ARAA:}
J.~E.~McClintock, R.~Shafee, R.~Narayan, R.~A.~Remillard, S.~W.~Davis and L.-X.~Li,	
``The Spin of the Near-Extreme Kerr Black Hole GRS 1915+105,''
Astro.\ Journal, {\bf 652}, 518-539 (2006),		
[arXiv:astro-ph/0606076].
	
\bibitem{Stu-Sla-Tor:2007a:ASTRA:}
	Z.~Stuchl{\'i}k, P.~Slan{\'y} and G.~T{\"o}r{\"o}k,
  ``Humpy LNRF-velocity profiles in accretion discs orbiting almost extreme Kerr black holes. A possible relation to quasi-periodic oscillations,''
	Astronom.\ Astrophys.\ {\bf 463}, 807-816 (2007).

\bibitem{Stu-Sla-Tor:2007b:ASTRA:}
Z.~Stuchl{\'i}k, P.~Slan{\'y} and G.~T{\"o}r{\"o}k,
``LNRF-velocity hump-induced oscillations of a Keplerian disc orbiting near-extreme Kerr black hole: a possible explanation of high-frequency QPOs in GRS 1915+105,''
Astronom.\ Astrophys.\ {\bf 470}, 401-404 (2007).
	
	
\bibitem{Asch:2004:ASTRA:}
B.~Aschenbach, ``Measuring mass and angular momentum of black holes with high-frequency quasi-periodic
oscillations,'' 
Astronomy and Astrophysics {\bf 425}, 1075 (2004),
[astro-ph/0406545].  
	
\bibitem{Stu-Sla-Tor-Abr:2005:PHYSR4:}
	Z.~Stuchl{\'i}k, P.~Slan{\'y}, G.~T{\"o}r{\"o}k and M.~A.~Abramowicz,
  ``Aschenbach effect: Unexpected topology changes in the motion of particles and fluids orbiting rapidly rotating Kerr black holes,''
	 Phys.\ Rev.\ D {\bf 71}, 024037  (2005),
	 [gr-qc/0411091].
	 
\bibitem{Stu-Kot-Tor:2013:ASTRA:}
  Z.~Stuchl{\'i}k, A.~Kotrlov{\'a} and G.~T{\"o}r{\"o}k,
  ``Multi-resonance orbital model of high-frequency quasi-periodic oscillations: possible high-precision determination of black hole and neutron star spin,''
	 Astronom.\ Astrophys.\ {\bf 552}, A10 (2013).
	
\bibitem{McCli-Rem:2011:CLAQG:}
 J.~E.~McClintock et all ,
``Measuring the spins of accreting black holes,''
Class.\ Quant.\ Grav.\  {\bf 28}, 114009 (2011),
[astro-ph.HE/1101.0811].	

\bibitem{Ali-etal:2013:CLAQG:}
A.~N.~Aliev, G.~D.~Esmer and P.~Talazan,
``Strong gravity effects of rotating black holes: quasi-periodic oscillations,''
Class.\ Quant.\ Grav.\  {\bf 30}, 045010 (2013),
[gr-qc/1205.2838].	
 
\bibitem{Ste-Gyu-Yaz:2013:PHYSR4:}
I.~Z.~Stefanov, G.~G.~Gyulchev and S.~S.~Yazadjiev,
``Quasiperiodic oscillations and Tomimatsu-Sato {$\delta$}=2 space-time,''
Phys.\ Rev.\ D {\bf 87}, 083005 (2013),
[astro-ph.HE/0411091].

\bibitem{Tor-etal:2011:ASTRA:}
	G.~T{\"o}r{\"o}k, A.~Kotrlov{\'a}, E.~{\v S}r{\'a}mkov{\'a} and Z.~Stuchl{\'i}k,
  ``Confronting the models of 3:2 quasiperiodic oscillations with the rapid spin of the microquasar GRS 1915+105,''
	Astronom.\ Astrophys.\ {\bf 531}, A59 (2011),
	[astro-ph.HE/1103.2438].

\bibitem{Wit:1985:NuclPhysB:}
   E.~Witten,
   ``Superconducting Strings,''
   Nucl.\ Phys.\ B {\bf 249}, 557 (1985).
   
\bibitem{Vil-She:1994:CSTD:}
  A.~Vilenkin and E.~P.~S.~Shellard, {\em Cosmic strings and other topological defects}, (Cambridge University Press, Cambridge 1994).

\bibitem{Bar:1973:BlaHol:}
  J. M. Bardeen, "Timelike and Null Geodesics in the Kerr Metrics."  from {\em Black Holes} (Eds.: C. D. Witt, B. S. D. Witt). Gordon and Breach, New York, London, Paris 1973, page 215.

\bibitem{Arnold:1978:book:} V.~I.~Arnold, {\em Mathematical methods of classical mechanics}, (Springer, New York, 1978).

\bibitem{Tor-Stu:2005:ASTRA:}
	G.~T{\"o}r{\"o}k and Z.~Stuchl{\'i}k,
  ``Radial and vertical epicyclic frequencies of Keplerian motion in the field of Kerr naked singularities. Comparison with the black hole case and possible instability of naked-singularity accretion discs,''
	 Astronom.\ Astrophys.\ {\bf 437}, 775-788 (2005).
	
\bibitem{Bar-Pre-Teu:1972:ApJ:}
J.~M.~Bardeen,  W.~H.~Press and S.~A.~Teukolsky,
``Rotating Black Holes: Locally Nonrotating Frames, Energy Extraction, and Scalar Synchrotron Radiation,''
Astro.\ Journal, {\bf 178}, 347-370 (1972) 	

\bibitem{Stu-Sche:2012:CLAQG:}
Z.~Stuchl{\'{\i}}k and J.~Schee,
``Observational phenomena related to primordial Kerr superspinars,''
Class.\ Quant.\ Grav.\  {\bf 29}, 065002 (2012).

\bibitem{Kot-Stu-Tor:2008:CLAQG:}
A.~Kotrlov{\'a},Z.~Stuchl{\'i}k and G.~T{\"o}r{\"o}k,
``Quasiperiodic oscillations in a strong gravitational field around neutron stars testing braneworld models,''
Class.\ Quant.\ Grav.\  {\bf 25}, 225016 (2008),
[gr-qc/0812.0720].	

\bibitem{Stu-Kot:2009:GenRelGrav:}
	Z.~Stuchl{\'i}k and A.~Kotrlov{\'a},
  ``Orbital resonances in discs around braneworld Kerr black holes,''
  Gen.\ Rel.\ Grav.\  {\bf 41}, 1305-1343 (2009),
  [astro-ph/0812.5066].	

\bibitem{Str:2001:ASTRJ2L:}
  T.~E.~Strohmayer, ``Discovery of a 450 Hz QPO from the Microquasar GRO J1655-40 with RXTE,''
  Astrophys. J. Lett. {\bf 552}, L49 (2001),
  [astro-ph/0104487].
    
\bibitem{Klu-Abr:2000:ASTROPH:}
  W.~Kluzniak and M.~A.~Abramowicz, ``The physics of kHz QPOs---strong gravity's coupled anharmonic oscillators,'' 
  ArXiv Astrophysics e-prints (2001) 
  [astro-ph/0104487].

\bibitem{Ali-Gal:1981:GENRG2:}
  A.~N.~Aliev nad D.~V.~Galtsov, ``Radiation from relativistic particles in non-geodesic motion in a strong gravitational field,'' Gen. Relativity Gravitation {\bf 13}, 899 (1981). 
  
\bibitem{Abr-etal:2004:ASTRJ2L:}
  M.~A.~Abramowicz, W.~Kluzniak, J.~E.~McClintock and R.~A.~Remillard, ``The Importance of Discovering
a 3 : 2 Twin-Peak Quasi-periodic Oscillation in an Ultraluminous X-Ray Source, or How to Solve the Puzzle of Intermediate-Mass Black Holes,''
  Astrophys. J. Lett. {\bf 609}, L63 (2004).   
  
\bibitem{Ter:2005:ASTRA:} 
	G.~T{\"o}r{\"o}k, ``A possible 3 : 2 orbital epicyclic resonance in QPOs frequencies of SgrA *,''
	Astronomy and Astrophysics, {\bf 440} 1 (2005),
	[astro-ph/0412500].

\bibitem{Lac-Cze-Abr:2006:astro-ph0607594:}
P.~Lachowicz, B.~Czerny and M.~A.~Abramowicz,
``Wavelet analysis of MCG-6-30-15 and NGC 4051: a possible discovery of QPOs in 2:1 and 3:2 resonance,''
ArXiv Astrophysics e-prints, {\bf 07} (Monthly Notices of the Royal Astronomical Society?) (2006),
[astro-ph/0607594].

\bibitem{Mid-Utt-Don:2011:}
M.~Middleton, P.~Uttley and C.~Done,
``Searching for the trigger of the active Galactic nucleus quasi-periodic oscillation: 8 years of RE J1034+396,''
Monthly Notices of the Royal Astronomical Society, {\bf 417} 250-260 (2011),
[astro-ph.CO/1106.1294].
	
\bibitem{Mid-Don-etal:2009:}
M.~Middleton, P.~Uttley and C.~Done,
``RE J1034+396: the origin of the soft X-ray excess and quasi-periodic oscillation,''
Monthly Notices of the Royal Astronomical Society, {\bf 394} 250-260 (2009),
[arXiv:0807.4847].

\bibitem{Nay-Moo:1979:NonOscilations:}
A.~H.~Nayfeh and D.~T.~Mook {\em Nonlinear Oscillations} (Wiley \& Sons, New York 1979)

\bibitem{Lan-Lif:1976:Mech:} L.~D.~Landau and E.~M.~Lifshitz, {\em Mechanics}, (Butterworth-Heinemann, Oxford 1976).

\bibitem{Stu-Kot-Tor:2011:ASTRA:}
Z.~Stuchl{\'{\i}}k, A.~Kotrlov{\'a} and G.~T{\"o}r{\"o}k, ``Resonant radii of kHz quasi-periodic oscillations in Keplerian discs orbiting neutron stars,''
Astronomy and Astrophysics, {\bf 525} A82 (2011),
[astro-ph.HE/1010.1951].

\end{thebibliography}
\end{document}